\documentclass[twocolumn]{aastex6}
\usepackage{amssymb}
\usepackage{amsmath}
\usepackage{amsfonts}
\usepackage{color}
\usepackage{graphicx}
\usepackage[varg]{txfonts}
\usepackage{natbib}
\usepackage{hyperref}
\usepackage{dcolumn}
\usepackage{bm}
\usepackage{lineno}
\shorttitle{Neutrino-driven explosion and SN~1999em}
\shortauthors{V. P. Utrobin et al.}
\begin{document}
%
\title{Light curve analysis of ordinary type IIP supernovae based on
   neutrino-driven explosion simulations in three dimensions}

\author{V.~P.~Utrobin\altaffilmark{1,2},
        A.~Wongwathanarat\altaffilmark{3,1},
        H.-Th.~Janka\altaffilmark{1}, and
        E.~M\"uller\altaffilmark{1}}

\altaffiltext{1}{
   Max-Planck-Institut f\"ur Astrophysik,
   Karl-Schwarzschild-Str. 1, 85748 Garching, Germany}
\altaffiltext{2}{
   State Scientific Center of the Russian Federation --
   Institute for Theoretical and Experimental Physics of
   National Research Center ``Kurchatov Institute'',
   B.~Cheremushkinskaya St. 25, 117218 Moscow, Russia; utrobin@itep.ru}
\altaffiltext{3}{
   RIKEN, Astrophysical Big Bang Laboratory, 2-1 Hirosawa, Wako,
   Saitama 351-0198, Japan}

\begin{abstract}
Type II-plateau supernovae (SNe IIP) are the most numerous subclass of
   core-collapse SNe originating from massive stars.
In the framework of the neutrino-driven explosion mechanism, we study the SN
   outburst properties for a red supergiant progenitor model and compare the
   corresponding light curves with observations of the ordinary Type
   IIP SN~1999em.
Three-dimensional (3D) simulations of (parametrically triggered) neutrino-driven explosions are
   performed with the (explicit, finite-volume, Eulerian, multifluid
   hydrodynamics) code {\sc Prometheus}, using a presupernova model of
   a 15\,$M_{\sun}$ star as initial data.
At approaching homologous expansion, the hydrodynamical and composition
   variables of the 3D models are mapped to a spherically symmetric
   configuration, and the simulations are continued with the (implicit,
   Lagrangian radiation-hydrodynamics) code {\sc Crab} to follow
   the blast-wave evolution during the SN outburst.
Our 3D neutrino-driven explosion model with an explosion energy of about
   $0.5\times10^{51}$\,erg produces $^{56}$Ni in rough agreement
   with the amount deduced from fitting the radioactively powered light-curve
   tail of SN~1999em.
The considered presupernova model, 3D explosion simulations, and light-curve  
   calculations can explain the basic observational features of SN~1999em,
   except for those connected to the presupernova structure of the outer
   stellar layers.
Our 3D simulations show that the distribution of $^{56}$Ni-rich matter
   in velocity space is asymmetric with a strong dipole component that
   is consistent with the observations of SN~1999em.
The monotonic luminosity decline from the plateau to the radioactive tail
   in ordinary SNe IIP is a manifestation of the intense turbulent mixing
   at the He/H composition interface.
\end{abstract}


\section{Introduction}
\label{sec:intro}
%
Massive stars in the range of $\sim$9$-25...30\,M_{\sun}$ produce
   a core of iron which collapses to a neutron star with the subsequent
   explosion ending the stellar lives as type II-plateau supernovae (SNe IIP)
   \citep[e.g.,][]{HFWLH_03}.
These hydrogen-rich objects are subdivided into the ordinary SNe IIP
   (e.g., SN~1999em, SN~2004et, SN~2012A), which show a plateau in the light
   curve, and the peculiar SNe IIP (e.g., SN~1987A, SN~2000cb, SN~2009E),
   which instead exhibit a dome-like light curve.
For a wealth of detail, insightful commentary, and further references, the
   reader may refer to large samples of SNe IIP collected by \citet{BH_09},
   \citet{AGH_14}, \citet{FPF_14}, and \citet{SSG_15}.
Relative fractions of the ordinary SNe IIP and the SN~1987A-like events are
   about 50\% \citep{LLC_11, SLFC_11} and 1--3\% \citep{PPN_12} of all
   core-collapse SNe (CCSNe), respectively.
It is theoretically established \citep{GIN_71, FA_77} and empirically
   confirmed \citep{Sma_09} that the most common ordinary SNe IIP originate
   from red supergiant (RSG) stars, while the peculiar objects are identified
   with the explosions of blue supergiant (BSG) stars \citep[e.g.,][]{ABKW_89}.

The phenomenon of CCSNe is very complex while considerable progress has been
   made in the last years in simulating these events \citep[see, e.g.,
   the reviews][]{JMS_16, Mul_16}, we need more observational information
   about how the explosion engine works.
Fortunately, very different energies and timescales inside the star just
   before the gravitational collapse of its central iron core allow us
   to divide the whole problem into two: an ``internal problem''
   (the gravitational collapse itself) and an ``external problem''
   (the collapse-initiated SN outburst) \citep{IN_89}, and to explore them
   independently.
In a previous paper we applied this approach to the well-observed and
   well-studied peculiar SN~1987A showing that the available pre-SN models,
   three-dimensional (3D) neutrino-driven explosion simulations with
   an approximate, parametrized neutrino engine (tuned to yield the observed
   values of the explosion energy and ejected $^{56}$Ni mass of SN~1987A),
   and light curve modeling can explain the basic observational data
   \citep{UWJM_15}.

\begin{figure}[t]
\centering
   \includegraphics[width=\hsize, clip, trim=16 159 71 101]{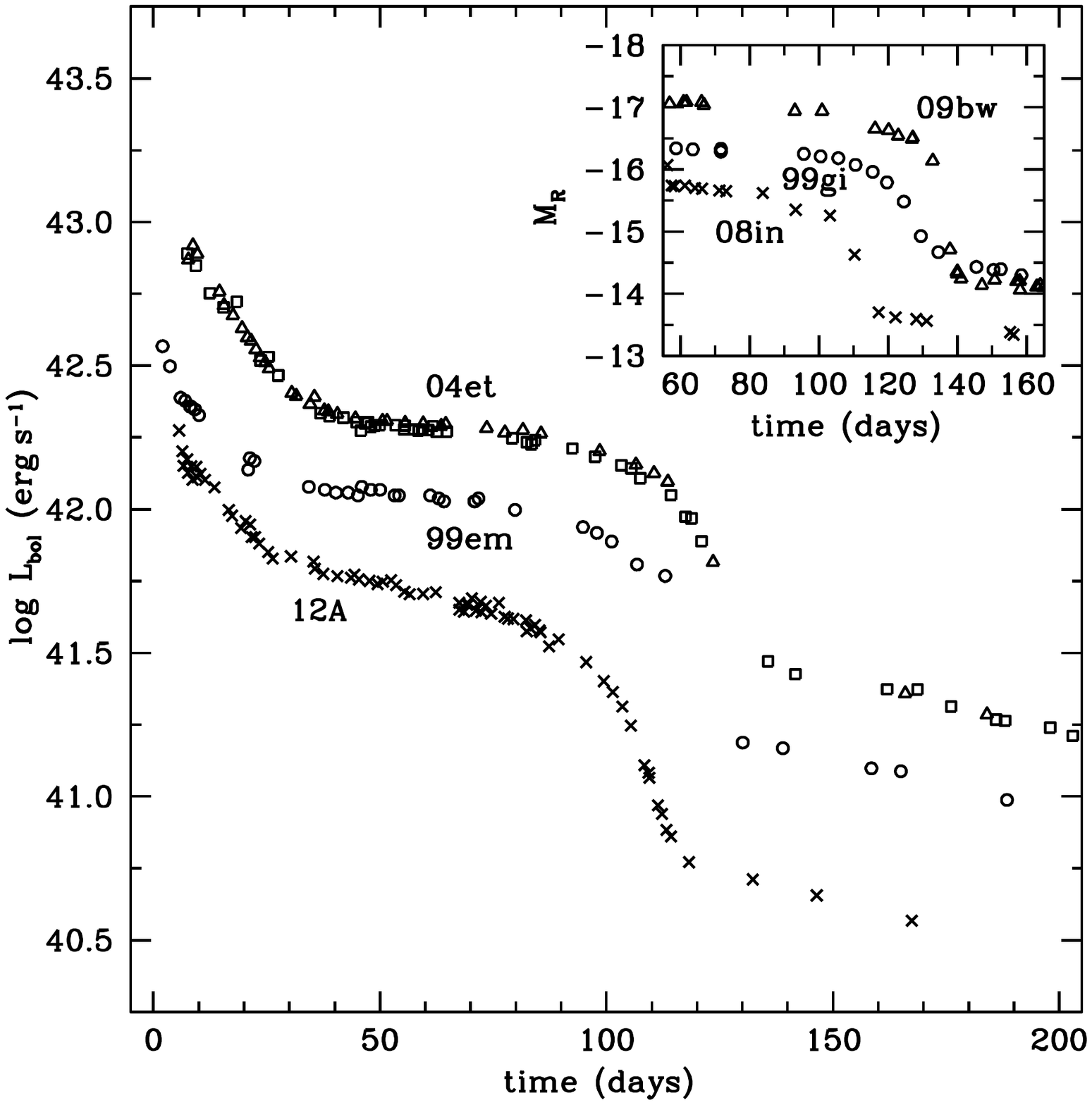}
   \caption{%
   Bolometric light curves of the luminous type IIP SN~2004et, evaluated
      from the $UBVRI$ magnitudes of \citet{SASM_06} (open triangles) and
      \citet{MDS_10} (open squares); the normal type IIP SN~1999em
      (open circles), estimated from the $UBVRI$ observations of \citet{ECP_03};
      and the sub-luminous type IIP SN~2012A (crosses), evaluated from
      the $UBVRIJHK$ magnitudes reported by \citet{TCF_13}.
   The inset shows the $R$ band light curves of the bright SN~2009bw
      (open triangles) \citep{ITP_12}, the normal SN~1999gi (open circles)
      \citep{LFL_02}, and the sub-luminous SN~2008in (crosses) \citep{RKB_11}.
   }
   \label{fig:obslc}
\end{figure}
In this work we continue the study of SNe IIP in the framework of the
   neutrino-driven explosion mechanism.
Ordinary SNe IIP exhibit a wide range of luminosities in the
   plateau phase and total masses of radioactive $^{56}$Ni, which is
   illustrated, for example, by the luminous SN~2004et, the normal SN~1999em,
   and the sub-luminous SN~2012A (Fig.~\ref{fig:obslc}).
All of these SNe originate from RSG stars, and their luminosities are produced
   by the release of the internal energy deposited during the shock wave
   propagation through the pre-SN envelope.
This energy release results in a monotonic decrease of the bolometric
   luminosity from the shock breakout to the radioactive tail.
Of particular interest is the luminosity decline at around 100\,days, which
   is related to the exhaustion of radiation energy from the inner layers
   of the ejecta.
These layers are subject to hydrodynamic instabilities and turbulent mixing
   at the C+O/He and He/H composition interfaces occurring during the
   explosion.
The $R$ band light curves of the bright SN~2009bw, the normal SN~1999gi, and
   the sub-luminous SN~2008in support the monotonicity of the luminosity
   decline (see inset in Fig.~\ref{fig:obslc}).

Important properties of the explosion can be deduced from observations
   and modeling of SNe IIP in the nebular phase when the ejecta become
   optically thin and nucleosynthesis products in the inner layers become
   visible.
Spectroscopic observations of the peculiar SN~1987A provide clear evidence
   for macroscopic mixing of the elements occurring during the explosion,
   which is well quantified \citep{UWJM_15}.
Unfortunately, among the ordinary SNe IIP there is no such well-observed and
   well-studied object as SN~1987A to firmly assess the extent of mixing of
   radioactive $^{56}$Ni and hydrogen in the ejecta.
\citet{MJS_12} found for all of the spectra in their sample of eight SNe IIP
   that the line profile shapes do not evolve with time.
They concluded that radioactive $^{56}$Ni is not concentrated in the central
   core of the ejecta, but instead is distributed by mixing to regions farther
   out.
\citet{ECP_03} analyzed the H$\alpha$ and He\,{\sc i} 10\,830~\AA\ lines in
   the normal SN~1999em at the nebular epoch and concluded that the
   distribution of the bulk of radioactive $^{56}$Ni can be approximated by
   a sphere of $^{56}$Ni with a velocity of 1500\,km\,s$^{-1}$, which is
   shifted towards the far hemisphere by about 400\,km\,s$^{-1}$.
\citet{Chu_07} interpreted the double-peak structure of H$\alpha$ at
   the nebular epoch in terms of asymmetric bipolar radioactive $^{56}$Ni
   jets.
\citet{MJS_12} also showed that the line profiles are intrinsically peaked
   in shape and suggested that mixing of the elements including hydrogen must
   occur in the ejecta to allow elements to be located at zero velocity.

Hydrodynamic models, solving the external problem of the SN explosion and
   based on evolutionary calculations of pre-SN models, are consistent
   with the observed light curves of ordinary SNe IIP only in basic aspects.
In particular, the theoretical light curves exhibit a conspicuous shoulder-like
   (\citealt[Fig.~5]{WH_07}; \citealt[Fig.~17]{MPR_15};
   \citealt[Fig.~33, lower panel]{SEWBJ_16}) or spike-like
   (\citealt[Fig.~11]{CDHLS_03}; \citealt[Fig.~1a]{You_04}) feature during
   the luminosity decline from the plateau to the radioactive tail, which
   is not observed (Fig.~\ref{fig:obslc}).
The feature occurs in hydrodynamic simulations of SN explosions triggered
   by both a thermal and/or kinetic bomb \citep{CDHLS_03, You_04, MPR_15}
   and a piston \citep{WH_07, SEWBJ_16}.
It is noteworthy that even invoking artificial mixing in the inner layers
   of the ejecta does not eliminate the unobserved post-plateau feature
   \citep{You_04}.

On the other hand, hydrodynamic simulations based on nonevolutionary pre-SN
   models fairly well fit the observations of the luminous SN~2004et
   \citep{UC_09}, the normal SN~1999em \citep{BBP_05, Utr_07}%
   \footnote{\citet{BBH_11} also used nonevolutionary pre-SN models
   to study SN~1999em, but invoked an extended $^{56}$Ni mixing to eliminate
   a bump feature at the end of the plateau and to obtain a nearly flat
   plateau of the light curve. See Sect.~\ref{sec:discssn} for details.},
   and the sub-luminous SN~2012A \citep{UC_15}.
In addition, the density profile of the nonevolutionary pre-SN models and
   the artificial mixing of radioactive $^{56}$Ni and hydrogen produce
   the monotonic luminosity decline from the plateau to the radioactive tail.

\begin{figure}[t]
\centering
   \includegraphics[width=\hsize, clip, trim=22 158 70 94]{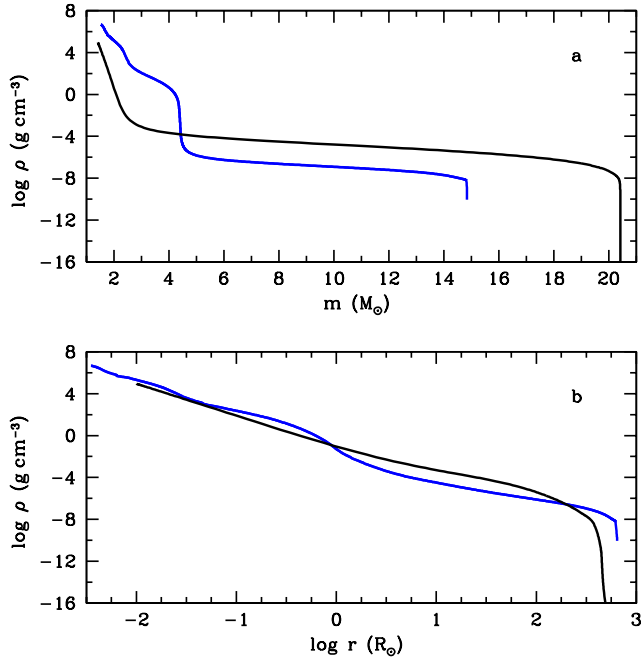}
   \caption{%
   Density profiles as functions of interior mass (panel \textbf{a})
      and radius (panel \textbf{b}) in pre-SN model L15 (blue line).
   The central collapsing core is omitted.
   For a comparison, the black line shows the structure of
       a nonevolutionary pre-SN model used by \citet{Utr_07}.
   }
   \label{fig:denmr}
\end{figure}
Our representative sample of ordinary SNe IIP, restricted to six different
   objects with luminosities in the range from low to high values, shows that
   SN~1999em is a typical one (Fig.~\ref{fig:obslc}).
For this reason, the normal SN~1999em is commonly considered as a template case
   of ordinary SNe IIP, and we will focus on it in our study.
As in the case of SN~1987A \citep{UWJM_15}, we carry out 3D hydrodynamic
   simulations of neutrino-driven explosions for the evolutionary pre-SN
   model of an RSG star.
These simulations yield a complex morphology of radioactive $^{56}$Ni
   and hydrogen mixing.
This morphology is retained in its global radial features after mapping to
   a spherically symmetric grid in order to simulate the evolution of
   SN~1999em after shock breakout and its light curve.
It is noteworthy that the turbulent mixing in the inner layers of the ejecta
   results in a modified density profile and sufficient mixing of both
   radioactive $^{56}$Ni and hydrogen such that the spike in the luminosity
   decline from the plateau to the radioactive tail nearly disappears
   (see below).

The paper is organized as follows.
In Sect.~\ref{sec:methmod} we briefly describe the pre-SN model,
   the 3D simulations of the neutrino-driven onset of the explosion,
   the 3D hydrodynamic modeling of the subsequent evolution until
   shock breakout, and the hydrodynamic light curve modeling.
We analyze the simulation results in Sect.~\ref{sec:results} and
   compare them with observations of SN~1999em in Sect.~\ref{sec:compobs}.
The origin of the unobserved luminosity spike in 1D hydrodynamic models is
   studied in Sect.~\ref{sec:orgnlsp}.
Finally, in Sect.~\ref{sec:discssn} we summarize and discuss our results.

\begin{figure}[t]
   \includegraphics[width=\hsize, clip, trim=17 158 72 321]{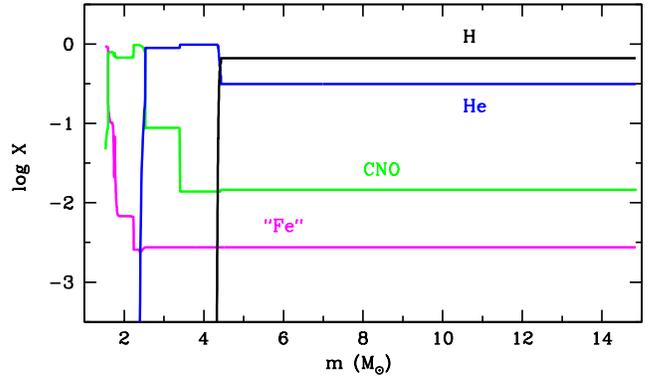}
   \caption{%
   Mass fractions of hydrogen (black line), helium
      (blue line), CNO-group elements (green line),
      and ``iron-group'' elements containing the heavy elements starting from
      silicon (magenta line) in the pre-SN model L15.
   }
   \label{fig:chcom}
\end{figure}
%
\section{Numerical methods and model overview}
\label{sec:methmod}
%
\subsection{Presupernova models}
\label{sec:methmod-psn}
%
\begin{deluxetable}{ l c c c c c c c c c }
\tabletypesize{\scriptsize}
\tablewidth{0pt}
\tablecaption{Presupernova models for red supergiants%
\label{tab:presnm}}
\tablehead{
\colhead{Model} & \colhead{$R_\mathrm{pSN}$}
                & \colhead{$M_\mathrm{He}^{\,\mathrm{core}}$}
                & \colhead{$M_\mathrm{pSN}$}
                & \colhead{$M_\mathrm{ZAMS}$}
                & \colhead{$X_\mathrm{surf}$}
                & \colhead{$Y_\mathrm{surf}$}
                & \colhead{$Z_\mathrm{surf}$}
                & \colhead{Ref.} \\
\colhead{} & \colhead{$(R_{\sun})$}
           & \colhead{$(M_{\sun})$}
           & \colhead{$(M_{\sun})$}
           & \colhead{$(M_{\sun})$}
           & \colhead{}
           & \colhead{}
           & \colhead{$(10^{-2})$}
           & \colhead{}
}
\startdata
 L15     & 627 & 4.35 & 14.85\tablenotemark{a} & 15.0\tablenotemark{b}
         & 0.666 & 0.314 & 2.0 & 1 \\
 Optimal & 500 & ---  & 20.60 & $\approx$22.2 & 0.735 & 0.248 & 1.7 & 2 \\
\enddata
\tablecomments{%
The columns give the name of the pre-SN model, its radius, $R_\mathrm{pSN}$;
   the helium-core mass, $M_\mathrm{He}^{\,\mathrm{core}}$;
   the pre-SN mass, $M_\mathrm{pSN}$; the progenitor mass,
   $M_\mathrm{ZAMS}$; the mass fraction of hydrogen, $X_\mathrm{surf}$;
   helium, $Y_\mathrm{surf}$; and heavy elements, $Z_\mathrm{surf}$,
   in the hydrogen-rich envelope at the stage of core collapse;
   and the corresponding reference.
}
\tablenotetext{a}{Actual mass of the pre-SN/progenitor model.}
\tablenotetext{b}{Nominal mass of the progenitor star.}
\tablerefs{
(1) \citet{LSC_00};
(2) \citet{Utr_07}.
}
\end{deluxetable}
We investigate a pre-SN model obtained for a 15\,$M_{\sun}$ progenitor star
   evolved by \citet{LSC_00} from the pre-main sequence up to the onset of
   collapse with the stellar evolution code {\sc Franec}
   (Table~\ref{tab:presnm}).
The model has solar chemical composition and was evolved with a nuclear
   reaction network extending up to $^{68}$Zn.
Convection was treated by means of the mixing-length formalism of
   B{\"o}hm-Vitense with the Schwarzschild criterion.
Semiconvection and overshooting were taken into account too.
The evolution was calculated without considering mass loss and the effects of
   rotation.
The pre-SN model, which we name L15, provides the initial data for our 3D
   neutrino-driven CCSN explosion simulations.
It has a helium core of 4.35\,$M_{\sun}$ and a radius of 627\,$R_{\sun}$
   typical of RSG stars (Table~\ref{tab:presnm}, Fig.~\ref{fig:denmr}).
The chemical composition of pre-SN model L15 is shown in
   Fig.~\ref{fig:chcom}.

For comparison, we provide the basic parameters of a nonevolutionary pre-SN
   model for SN~1999em (optimal model in Table~\ref{tab:presnm}) constructed
   by \citet{Utr_07}.
Creating an adequate hydrodynamic explosion model is based on the method of
   trial and error using general relations between its basic physical
   parameters, the density and chemical composition distributions of
   the pre-SN model, and observed properties \citep{Utr_07}.
A comprehensive comparison of the calculated observable model properties with
   photometric and spectroscopic observations allows one to select the optimal
   hydrodynamic model.
Applying this procedure to SN~1999em results in a model that is quite
   different from that of the evolutionary pre-SN model L15.
In particular, the nonevolutionary model has no steep density gradients like
   those present in the pre-SN model L15 (Fig.~\ref{fig:denmr}).
We note that such a nonevolutionary pre-SN model mimics the intense
   turbulent mixing occurring during the explosion at the locations of the
   Si/O, (C+O)/He, and He/H composition interfaces \citep{UC_08}.
In addition, a sharp density decline in the outermost layers of the model is
   favorable for an acceleration of these layers to the high velocities
   inferred from the spectral lines of SN~1999em.

\subsection{3D hydrodynamic modeling until shock breakout}
\label{sec:methmod-3Dmod}
%
Our 3D calculations are performed with the explicit finite-volume, Eulerian,
   multifluid hydrodynamics code {\sc Prometheus} \citep{FAM_91, MFA_91a,
   MFA_91b}.
It integrates the multidimensional hydrodynamic equations using dimensional
   splitting \citep{Str_68}, piecewise parabolic reconstruction \citep{CW_84},
   and a Riemann solver for real gases \citep{CG_85}.
To relax the restrictive CFL-timestep condition and to avoid numerical
   artifacts near the polar axis, {\sc Prometheus} employs an axis-free
   overlapping ``Yin-Yang'' grid \citep{KS_04} in spherical polar coordinates,
   which was implemented into the code by \citet{WHM_10}.
Newtonian self-gravity is taken into account by solving Poisson's equation
   in its integral form, using an expansion into spherical harmonics
   \citep{MS_95}.

The SN explosion is triggered by imposing a suitable value of the neutrino
   luminosities at an inner grid boundary located at an enclosed mass of
   1.1\,$M_{\sun}$ well inside of the neutrinosphere.
Outside this boundary, which is moved to mimic the contracting proto-neutron
   star, we apply an approximate neutrino transport and neutrino-matter
   interactions as described in \citet{SKJM_06}.
The explosion energy of the model is determined by the imposed isotropic
   neutrino luminosity, whose temporal evolution we prescribe too, and
   the accretion luminosity which results from the progenitor dependent mass
   accretion rate and the gravitational potential of the contracting neutron
   star.  
To follow the explosive nucleosynthesis approximately, a small $\alpha$-chain
   reaction network, similar to the network described in \citet{KPSJM_03},
   is solved.

\begin{deluxetable}{ l c c c c c c c l }
\tabletypesize{\scriptsize}
\tablewidth{0pt}
\tablecaption{Basic properties of averaged 3D explosion models and 1D piston
   simulations%
\label{tab:3Dsim}}
\tablehead{
\colhead{Model} & \colhead{$M_\mathrm{CC}$}
                & \colhead{$M_\mathrm{env}$}
                & \colhead{$E_\mathrm{exp}$}
       & \colhead{$M_\mathrm{Ni}^{\,\mathrm{min}}$}
       & \colhead{$M_\mathrm{Ni}^{\,\mathrm{max}}$}
       & \colhead{$M_\mathrm{Ni}$}
       & \colhead{$t_\mathrm{map}$}
       & \colhead{Remark} \\
\colhead{} & \multicolumn{2}{c}{$(M_{\sun})$}
       & \colhead{(B)}
       & \multicolumn{3}{c}{$(10^{-2}\,M_{\sun})$}
       & \colhead{($10^{5}$\,s)}
       & \colhead{}
}
\startdata
L15-le  & 2.04 & 12.81 & 0.54 & 1.8 & 5.7 & 3.6 & 1.11 & 3D Sim. \\
L15-he  & 1.79 & 13.06 & 0.93 & 2.8 & 9.6 & 6.8 & 0.88 & 3D Sim. \\
L15-lm  & 2.04 & 11.16 & 0.54 & 1.8 & 5.7 & 3.6 & 0.68 & 3D Sim. \\
L15-pn  & 1.53 & 13.32 & 0.50 & --- & --- & 3.6 & ---  & Piston \\
Optimal & 1.60 & 19.00 & 1.30 & --- & --- & 3.6 & ---  & Piston \\
\enddata
\tablecomments{%
The computed models are based on the corresponding pre-SN models
   of Table~\ref{tab:presnm}.
$M_\mathrm{CC}$ is the mass of the collapsed remnant;
   $M_\mathrm{env}$ the ejecta mass;
   $E_\mathrm{exp}$ the explosion energy;
   $M_\mathrm{Ni}^{\,\mathrm{min}}$ the mass of radioactive $^{56}$Ni produced
   directly by our $\alpha$-chain reaction network;
   $M_\mathrm{Ni}^{\,\mathrm{max}}$ the aggregate mass of directly produced
   $^{56}$Ni and tracer nucleus; and
   $M_\mathrm{Ni}$ the radioactive $^{56}$Ni mass used in the 1D simulations.
$t_\mathrm{map}$ is the time at which the 3D simulations are mapped to
   a spherically symmetric grid.
}
\end{deluxetable}
As our reference model we choose a 3D CCSN explosion model calculated by
   \citet{WJM_13}, which was denoted L15-5 in \citet{WJM_13}.
It is based on the pre-SN model L15 (Table~\ref{tab:presnm}) and was evolved
   until 1.4\,s after core bounce with the simplified, gray neutrino transport.
We simulated the subsequent evolution until shock breakout for two models,
   L15-le and L15-he, which differ mainly by their explosion energies
   (Table~\ref{tab:3Dsim}).
To increase the explosion energy without destroying the global morphology of
   the ejecta, we boosted model L15-he by a constant neutrino-driven
   wind at the inner grid boundary for another 2\,s, whereas we simulated
   model L15-le with a neutrino-driven wind that declines with time by
   a power law \citep{WMJ_15}.
This treatment is supposed to single out the dependence of the light curve
   modeling on the explosion energy, while preserving the explosion asymmetry
   of the 3D reference model at 1.4\,s.
Redoing a 3D explosion simulation from core bounce on with a final explosion
   energy similar to model L15-he but different (stochastically developing)
   ejecta morphology, we do neither expect significant differences in the
   $^{56}$Ni yield (explosion energy and $^{56}$Ni production are tightly
   correlated in neutrino-driven explosions; see \citet{SEWBJ_16}) nor in
   the extent of $^{56}$Ni mixing, which depends mostly on the explosion
   energy and the progenitor structure \citep{WMJ_15}.
For explosion simulations with energies similar to model L15-he we therefore
   expect very similar light curves.

Basic properties of the averaged 3D simulations for two computed models L15-le
   and L15-he are listed in Table~\ref{tab:3Dsim}.
The explosion energy $E_\mathrm{exp}$ is defined as the sum of the total (i.e.,
   internal plus kinetic plus gravitational) energy of all grid cells
   at the time when the 3D data are mapped to the 1D grid for simulating
   the light-curve formation.
Throughout this paper, we employ the energy unit
   $1\,\mathrm{bethe}=1\,\mathrm{B}=10^{51}$\,erg.

Because the approximate treatment of neutrino-transport employed in the 3D
   explosion model does not allow us to capture accurately the time evolution
   of the electron fraction in the neutrino-processed ejecta, where
   a significant fraction of $^{56}$Ni can be produced, we provide minimum
   and maximum $^{56}$Ni yields of the 3D models in Table~\ref{tab:3Dsim}.
The minimum yield of $^{56}$Ni is produced directly (mostly in shock-heated
   ejecta) by our $\alpha$-chain reaction network.
The maximum yield of $^{56}$Ni, in turn, is given by the mass of $^{56}$Ni plus
   the mass of a tracer species which is produced in neutrino-heated ejecta
   under conditions of neutron excess.

To match the duration of the plateau phase of the light curve of SN~1999em,
   we construct an additional model, L15-lm in Table~\ref{tab:3Dsim}, which
   is based on model L15-le, but whose ejecta mass is
   decreased by removing the outermost hydrogen-rich layers of
   $1.65\,M_{\sun}$.
This reduction in the ejecta mass can be interpreted as a mass loss of
   $1.65\,M_{\sun}$ by the progenitor star (Table~\ref{tab:presnm}).
A justification of this assumption can be obtained from an analysis of
   the observations of SN~1999em, in which \citet{CCU_07} showed that its
   progenitor star had lost about 1\,$M_{\sun}$ during the RSG stage.
For a 20\,$M_{\sun}$ star with a luminosity of $\sim$10$^{5}$\,$L_{\sun}$ and
   a mass-loss rate of less than 10$^{-7}$\,$M_{\sun}$\,year$^{-1}$
   \citep{Krt_14}, the mass lost by winds is less than 1\,$M_{\sun}$ during
   the $\sim$10$^{7}$\,year it spends on the main sequence.
Thus, the total mass lost by the progenitor of SN~1999em does not exceed
   2\,$M_{\sun}$.

To compare the hydrodynamic models based on 3D neutrino-driven explosion
   simulations and the evolutionary pre-SN model L15 with 1D explosions
   triggered by a piston, we compute model L15-pn using
   the evolutionary pre-SN model L15, and also consider the optimal model based
   on the nonevolutionary pre-SN model \citep{Utr_07} (Table~\ref{tab:3Dsim}).
In 3D explosion simulations $^{56}$Ni mixing results from hydrodynamic
   instabilities.
In contrast, both in model L15-pn and the optimal model, which are exploded
   with a 1D piston, $^{56}$Ni is mixed artificially and nearly uniformly in
   velocity space up to $\approx$450 and 660\,km\,s$^{-1}$, respectively.

\subsection{Mapping 3D simulations to 1D problem}
\label{sec:methmod-mapping}
%
\begin{figure*}[ht]
\centering
   \includegraphics[width=0.30\hsize, clip, trim=60 150 90 100]{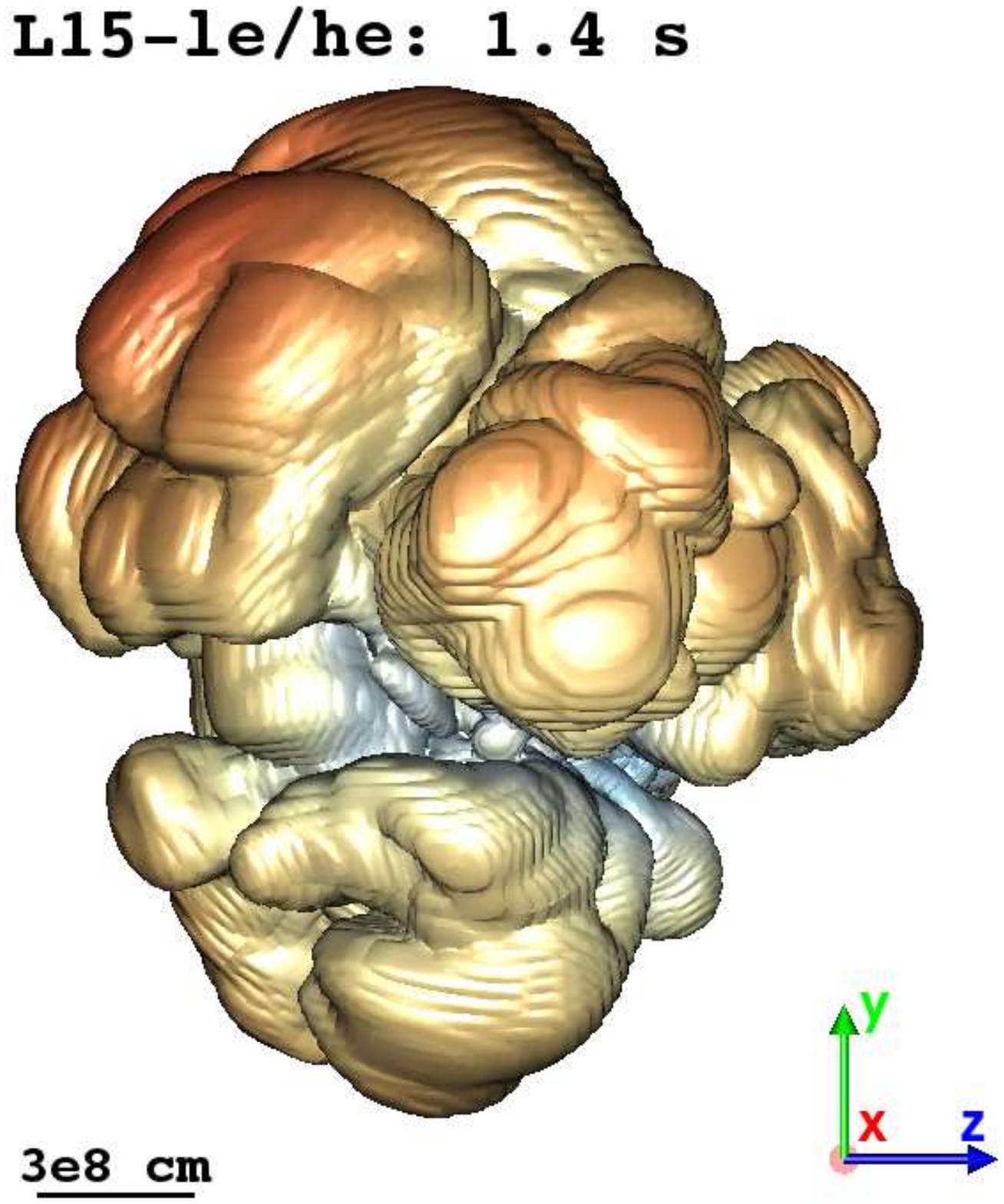}
   \includegraphics[width=0.30\hsize, clip, trim=60 150 90 100]{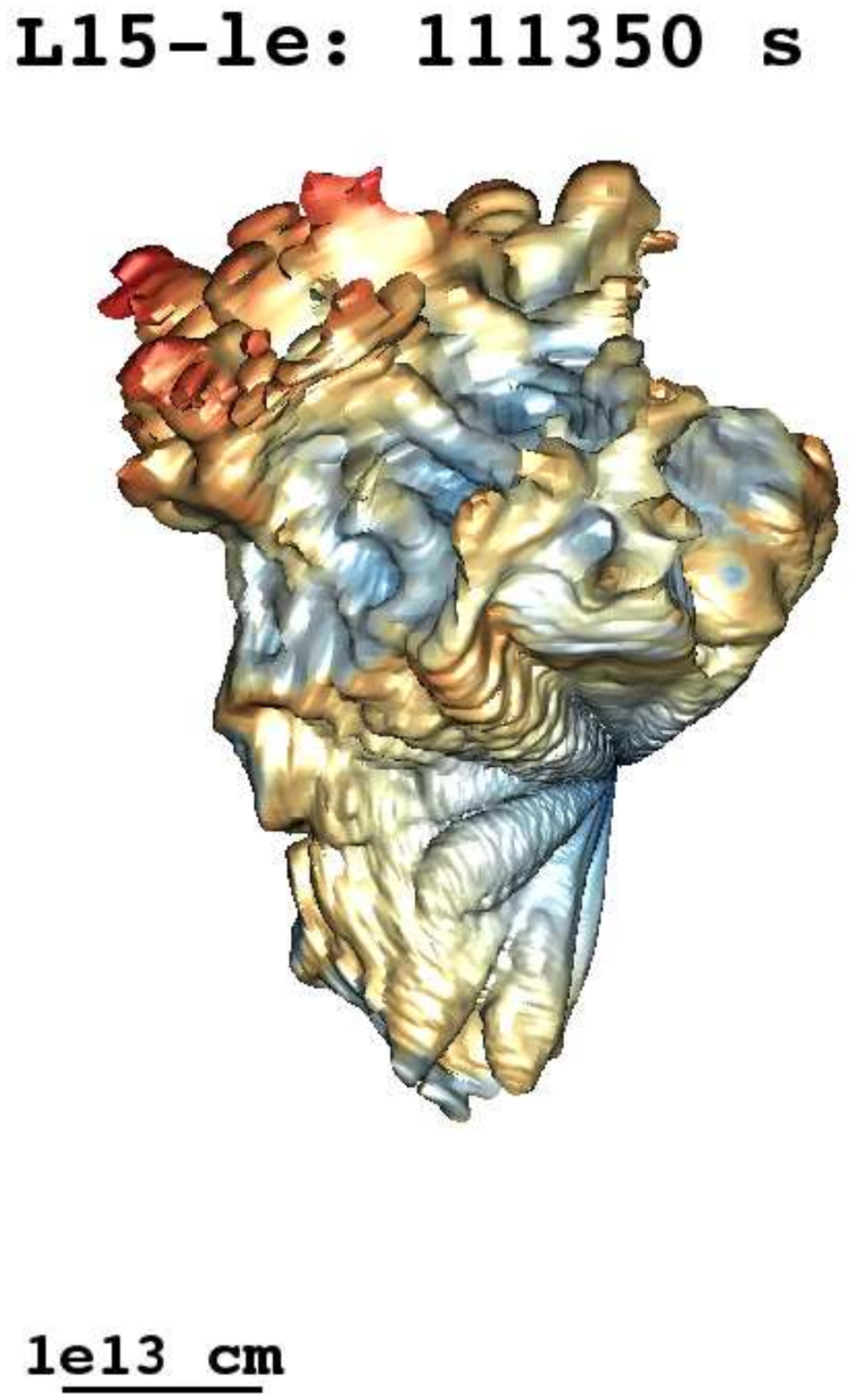}
   \includegraphics[width=0.30\hsize, clip, trim=60 155 90 100]{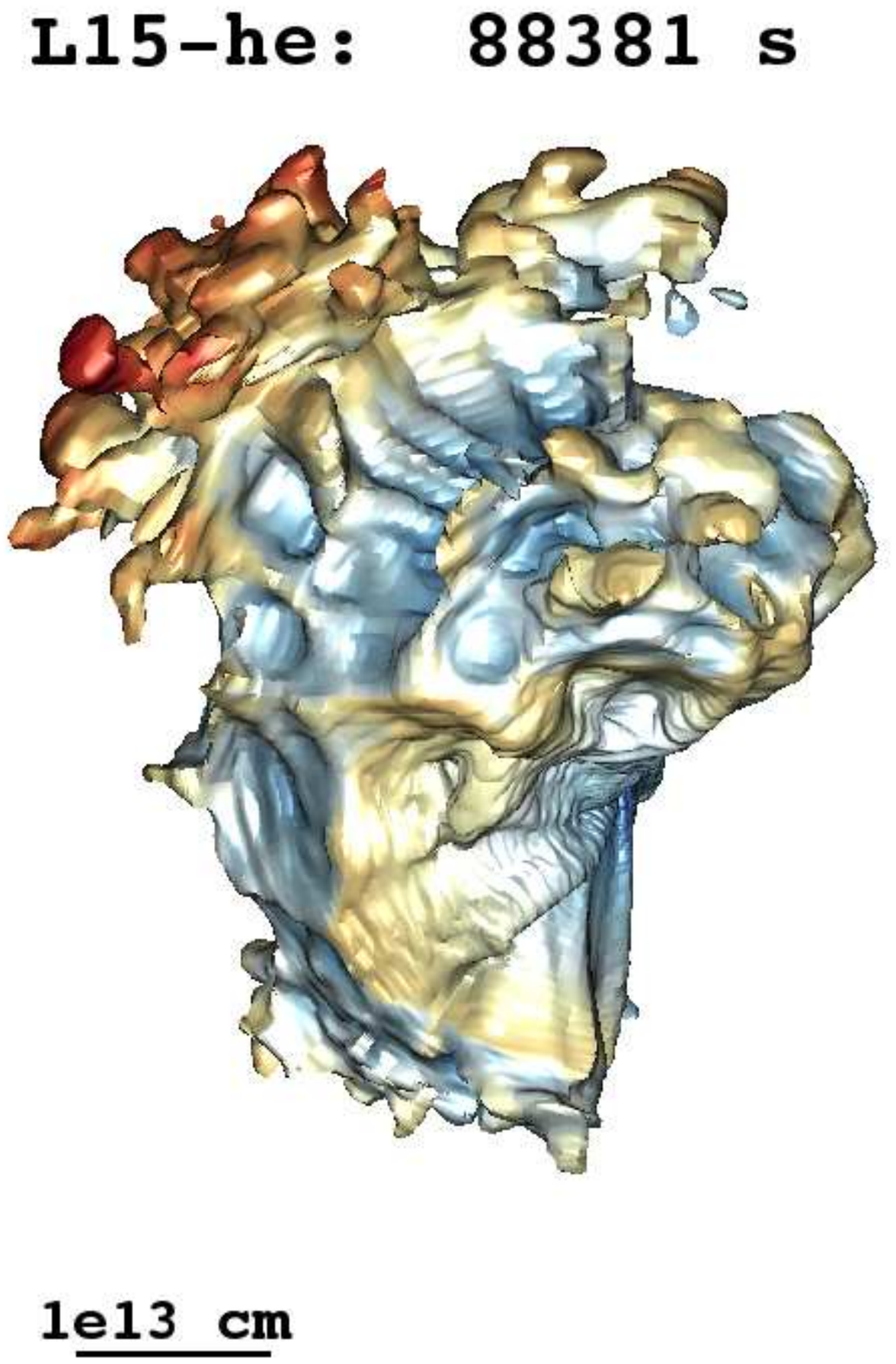}\\
   \includegraphics[width=0.30\hsize, clip, trim=60 80 90 210]{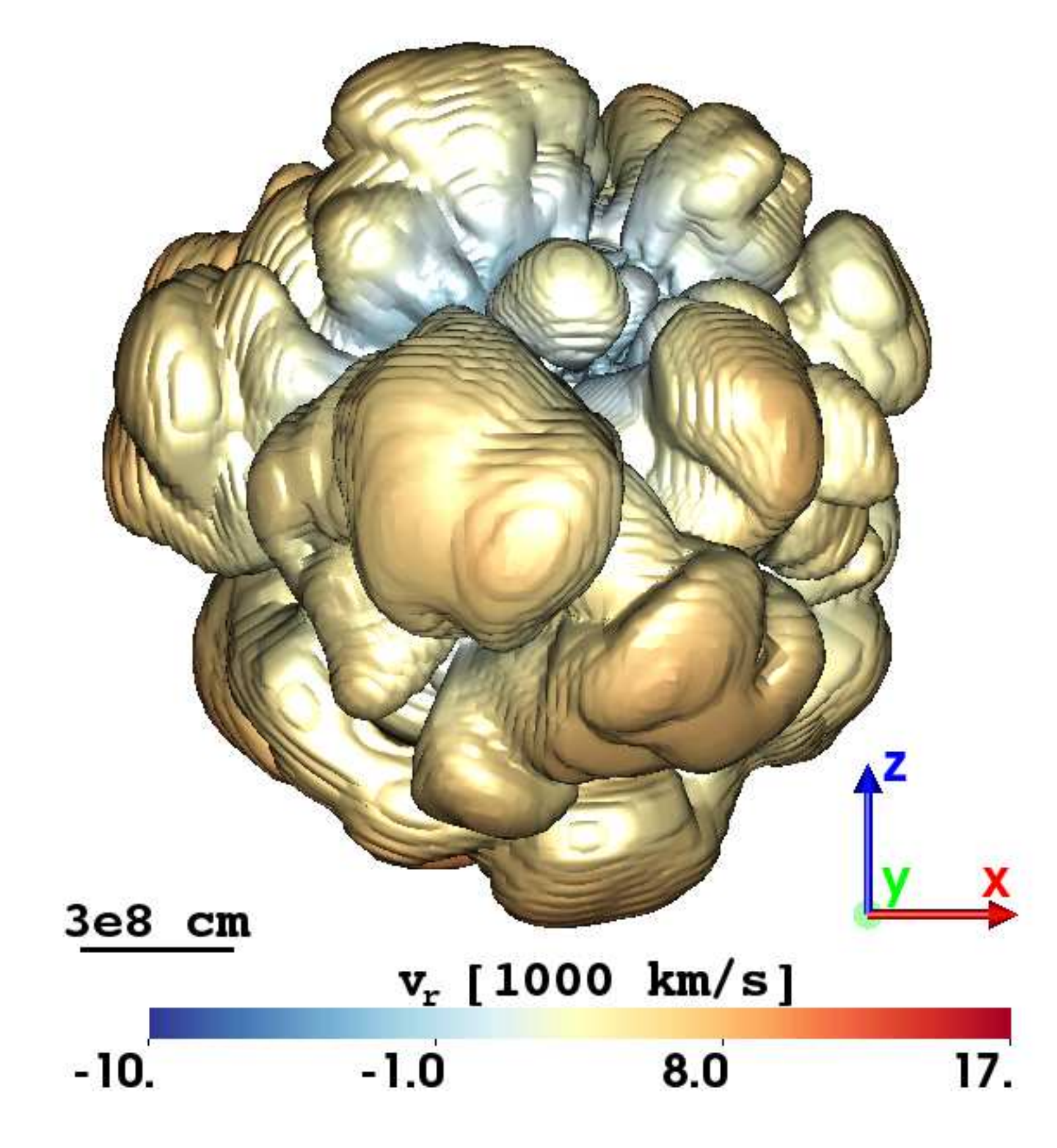}
   \includegraphics[width=0.30\hsize, clip, trim=60 80 90 210]{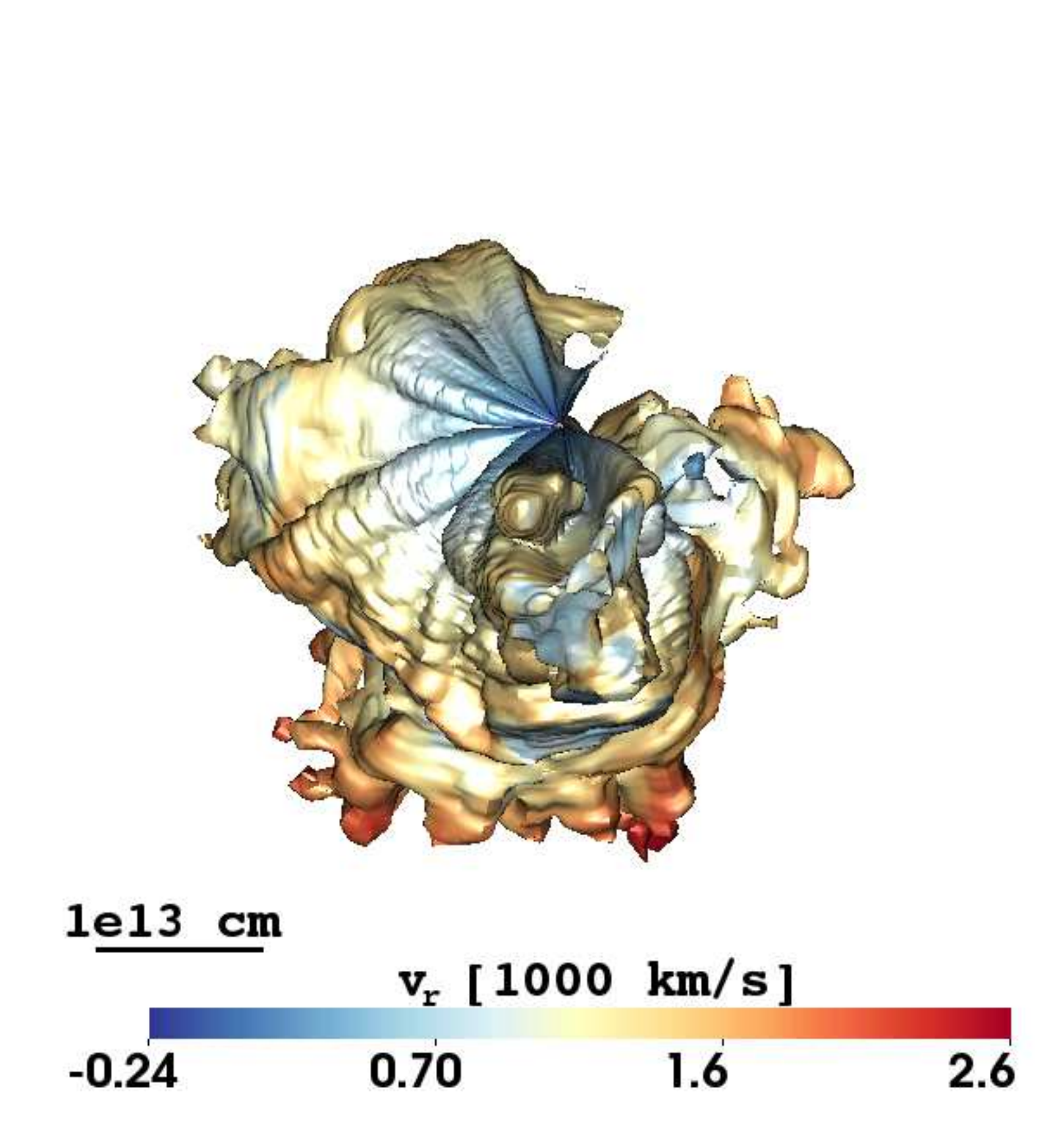}
   \includegraphics[width=0.30\hsize, clip, trim=60 80 90 210]{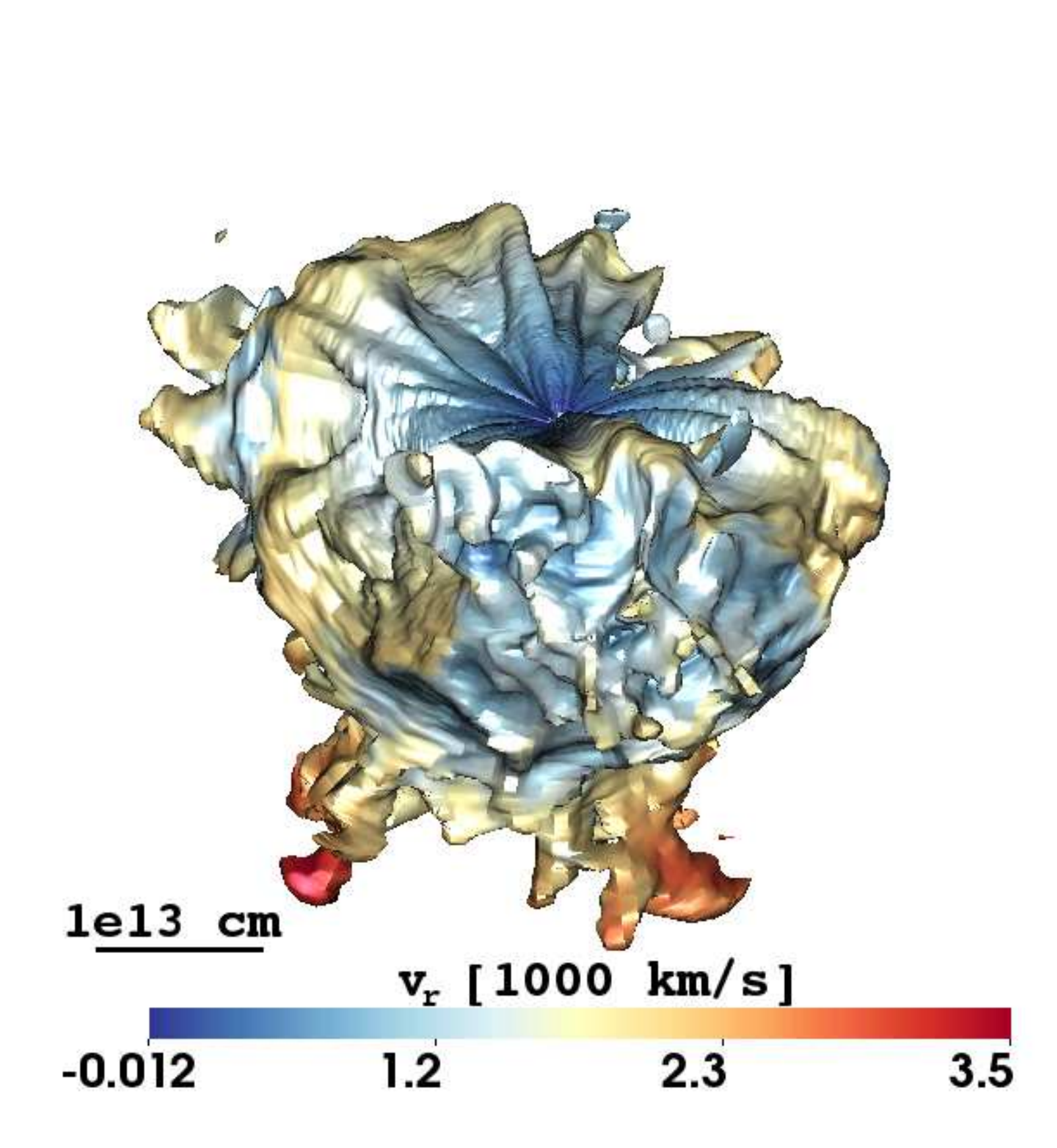}\\
   \caption{
   Morphology of radioactive $^{56}$Ni-rich matter produced by explosive
      burning in shock-heated ejecta.
   The snapshots display isosurfaces where the mass fraction of $^{56}$Ni
      (plus the neutron-rich tracer nucleus originating from matter with
      neutron excess) equals $3\%$.
   The isosurfaces are shown for 3D models L15-le and L15-he at two different
      epochs: at $t=1.4$\,s after bounce, i.e.\ before the SN shock crosses
      the C+O/He composition interface in the progenitor star,
      (\emph{left panels\/}) and before the shock breakout from the stellar
      surface at the mapping epochs of $t=111\,350$\,s (\emph{middle panels\/})
      and $88\,381$\,s (\emph{right panels\/}), respectively.
   The \emph{upper row} shows snapshots viewed along the $x$-direction,
      while the \emph{lower row} displays snapshots viewed along
      the $y$-direction.
   The colors give the radial velocity on the isosurface, the color coding
      being defined at the bottom of the lower row.
   At the top of each panel in the upper row, we give the name of the model
      and the post-bounce time of the snapshot.
   The size of the displayed volume and of the asymmetric structures can be
       estimated from the yardsticks given in the lower left corner of each
       panel.
   One notices that the final asymmetry of the $^{56}$Ni-rich ejecta
      in (velocity) space exhibits a pronounced dipole component.
   }
   \label{fig:3D_models}
\end{figure*}
To follow the evolution beyond shock breakout, we average the 3D hydrodynamic
   flow and the distribution of chemical elements on a spherically symmetric
   grid at chosen times and interpolate them onto the Lagrangian (mass) grid
   of the 1D simulations.
These data are used as the initial conditions for the external problem of
   the hydrodynamic modeling of the SN outburst.
With the hydrodynamic flow being given by our 3D simulations of neutrino-driven
   explosions, there is no need to initiate the explosion by a supersonic
   piston.
Another way of triggering a SN explosion is a thermal and/or kinetic bomb,
   whose effect on the light curve from the shock breakout to the end of
   the plateau is indistinguishable from that of piston-driven explosion,
   if the bomb does not extend significantly beyond the Si shell and
   acts no longer than a few seconds.

\subsection{Light curve modeling}
\label{sec:methmod-lcurves}
%
We simulate the evolution of the SN outburst after shock breakout with the
   implicit Lagrangian radiation hydrodynamics code {\sc Crab} \citep{Utr_04,
   Utr_07}.
It solves the set of spherically symmetric hydrodynamic equations including
   self-gravity, and a radiation transfer equation in gray approximation
   \citep[e.g.,][]{MM_84}.
The time-dependent radiative transfer equation, written in a comoving
   frame of reference to accuracy of order $v/c$ ($v$ is the fluid velocity,
   $c$ is the speed of light), is solved using the zeroth and first angular
   moments of the nonequilibrium radiation intensity.
This system of moment equations is closed by calculating a variable Eddington
   factor directly taking into account scattering of radiation in the SN
   ejecta.
The diffusion of equilibrium radiation, occurring in the inner, optically
   thick layers of the ejecta, is treated in the approximation of radiative
   heat conduction.
The resultant set of equations is discretized spatially using the method of
   lines \citep[e.g.,][]{HNW_93, HW_96}.
Energy deposition of gamma rays with energies of about 1\,MeV from the
   decay chain $^{56}$Ni $\to ^{56}$Co $\to ^{56}$Fe is calculated by solving
   the corresponding gamma-ray transport.
The equation of state, the mean opacities, and the thermal emission coefficient
   are computed taking non-LTE and non-thermal effects into account.
In addition, the contribution of spectral lines to the opacity in a medium
   expanding with a velocity gradient is estimated by the generalized formula
   of \citet{CAK_75}.
We refer to \citet{UWJM_15} and references therein for details on the numerical
   set up.

\section{Results}
\label{sec:results}
%
\subsection{Mixing in 3D explosion simulations}
\label{sec:results-3Dexp}
%
First of all we outline the development of neutrino-driven explosions after
   core bounce \citep[see, e.g.,][for details]{WMJ_15}.
As an illustrative example, we consider our fiducial model L15-le
   (Table~\ref{tab:3Dsim}).
Growth of Rayleigh-Taylor mushrooms from the imposed seed perturbations are
   first visible at about $t=80$\,ms after bounce.
These small mushrooms merge into high-entropy bubbles, which rise outward to
   the immediate vicinity behind the SN shock, providing pressure support
   for the SN shock against the ram pressure of infalling material.
Supported by convective overturn and global shock motions due to the standing
   accretion shock instability \citep[SASI; e.g.,][]{BMD_03, SJFK_08, OKY_06},
   the delayed, neutrino-driven explosion sets in at roughly $t=516$\,ms after
   bounce.

Neutrino-driven convection and SASI mass motions during the launch of
   the explosion create the morphology of the neutrino-heated ejecta.
\citet{KPSJM_03} showed that $^{56}$Ni is explosively produced in ``pockets''
   between the high-entropy bubbles of neutrino-heated matter, i.e.\ the
   distribution of $^{56}$Ni reflects the asymmetries of the first second of
   the explosion (Fig.~\ref{fig:3D_models}, left panels).
Around $t=2.47$\,s after bounce, when the SN shock reaches the (C+O)/He
   composition interface, the morphology of the $^{56}$Ni-rich ejecta still
   resembles their initial asymmetries.

The further evolution of the explosion depends strongly on the density profile
   of the pre-SN star, because the profile determines the amount of
   deceleration that the shock experiences while propagating through the helium
   and hydrogen layers of the star.
Dense shells form behind the decelerating shock, which become Rayleigh-Taylor
   unstable.
Thereby the initial morphology of the metal (iron) rich ejecta can be modified,
   if the dense shells form before these ejecta reach the formation sites
   of the dense shells. 
Whether this happens depends on the relative speed of the shock wave and
   the innermost ejecta \citep{WMJ_15}.

Figure~\ref{fig:3D_models} (middle and left panels) shows that
   the morphology of the $^{56}$Ni-rich ejecta of model L15-le at late times
   reflects the initial asymmetries of the neutrino-heated bubbles,
   i.e.\ global asymmetry is imprinted by the explosion mechanism rather than being
   a result of (secondary) Rayleigh-Taylor instabilities at the pre-SN
   composition interfaces.
The figure also shows (middle panels) that the $^{56}$Ni-rich ejecta have
   a pronounced dipolar component of asymmetry, the radial velocities of the
   fastest $^{56}$Ni-rich fingers reaching $\approx$2600\,km\,s$^{-1}$.
Relative to the center of mass motion of the $^{56}$Ni ejecta
   (containing the directly produced $^{56}$Ni and 50\% of the tracer 
   nucleus) the velocity asymmetry amounts to nearly 300\,km\,s$^{-1}$.

Models L15-le and L15-he have significantly different explosion energies
   (0.54\,B and 0.93\,B; see Table~\ref{tab:3Dsim}).
The more energetic model L15-he is evolving faster than model L15-le, 
   i.e.\ the velocities of its shock and $^{56}$Ni-rich matter are 
   higher than those of the less energetic model L15-le.
However, because both models are based on the same explosion calculation
   (L15 at 1.4\,s post bounce) and the increased explosion energy of
   model L15-he is the result of a subsequently imposed spherically 
   symmetric constant neutrino-driven wind, we find no significant 
   differences between the morphologies of their $^{56}$Ni-rich ejecta 
   (Fig.~\ref{fig:3D_models}, middle and right panels).

The asymmetry of $^{56}$Ni-rich matter with a strong dipole component in
   model L15-le differs considerably from that of the more energetic
   models L15-1-cw and L15-2-cw, which are based on the same pre-SN star
   but have an explosion energy of 1.75\,B and 2.75\,B, respectively
   \citep{WMJ_15}.
The model sequence L15-le, L15-1-cw, and L15-2-cw, along which the explosion
   energy increases, reveals a clear correlation between the morphology of
   $^{56}$Ni-rich matter and the explosion energy.
These models demonstrate that long-lasting phases of SASI activity tend
   to give rise to more extreme asymmetries of the distribution of
   $^{56}$Ni-rich ejecta, correlated with lower explosion energies,
   in which case low-mode asymmetries have more time to grow during shock
   revival before the explosion sets in.

As the main shock propagates through the exploding star until it finally breaks
   out from the stellar surface, the initial explosion asymmetry and shock
   deformation shape the ejecta and thus determine the global morphology also
   of the outer layers.
To measure the asphericity of the outer layers in the 3D simulations of model
   L15-le, we approximated the photosphere by an ellipsoidal surface for
   a density close to the photospheric density found in the averaged
   3D simulations.
The maximum ratio of the semiaxes for the approximation thus obtained at
   1.86\,days is 1.153.

\subsection{Approach to homologous expansion}
\label{sec:results-ahexp}
\begin{figure}[t]
   \includegraphics[width=\hsize, clip, trim=59 158 44 300]{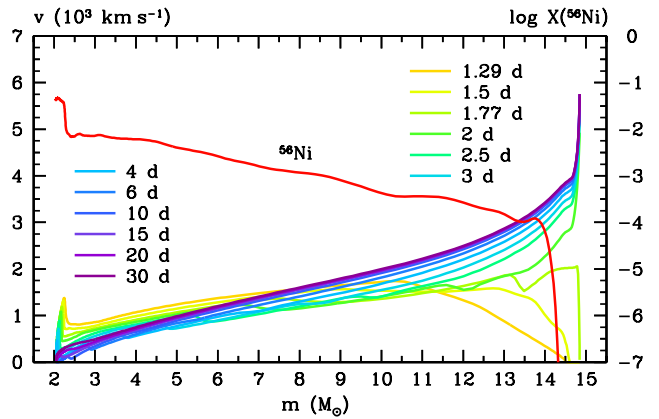}
   \caption{%
   The approach to homologous expansion is shown by comparing the velocity
      profiles of model L15-le at different moments in the 1D simulations.
   The red line gives the distribution of the $^{56}$Ni mass fraction at the
      mapping epoch (Table~\ref{tab:3Dsim}).
   }
   \label{fig:aphexp}
\end{figure}

%
Homologous expansion occurs when the contributions of pressure gradients and
   gravitational forces to the momentum equation may be neglected.
We show the approach of the flow to homology for model L15-le in
   Fig.~\ref{fig:aphexp}, which covers the evolution of the (Lagrangian)
   velocity profile from the time of mapping from 3D to 1D at 1.29\,days
   up to 30\,days.
At the mapping epoch, well before the time of shock breakout at 1.77\,days,
   the hydrodynamic flow is far from homologous, because the velocity profile
   of the outer layers evolves still significantly in the 1D simulations.
This implies that in the 3D neutrino-driven simulations outward mixing of
   radioactive $^{56}$Ni and inward mixing of hydrogen-rich matter in velocity
   space will continue until complete homology is reached.

\begin{figure*}[t]
\centering
   \includegraphics[width=0.9\hsize, clip, trim=32 173 70 202]{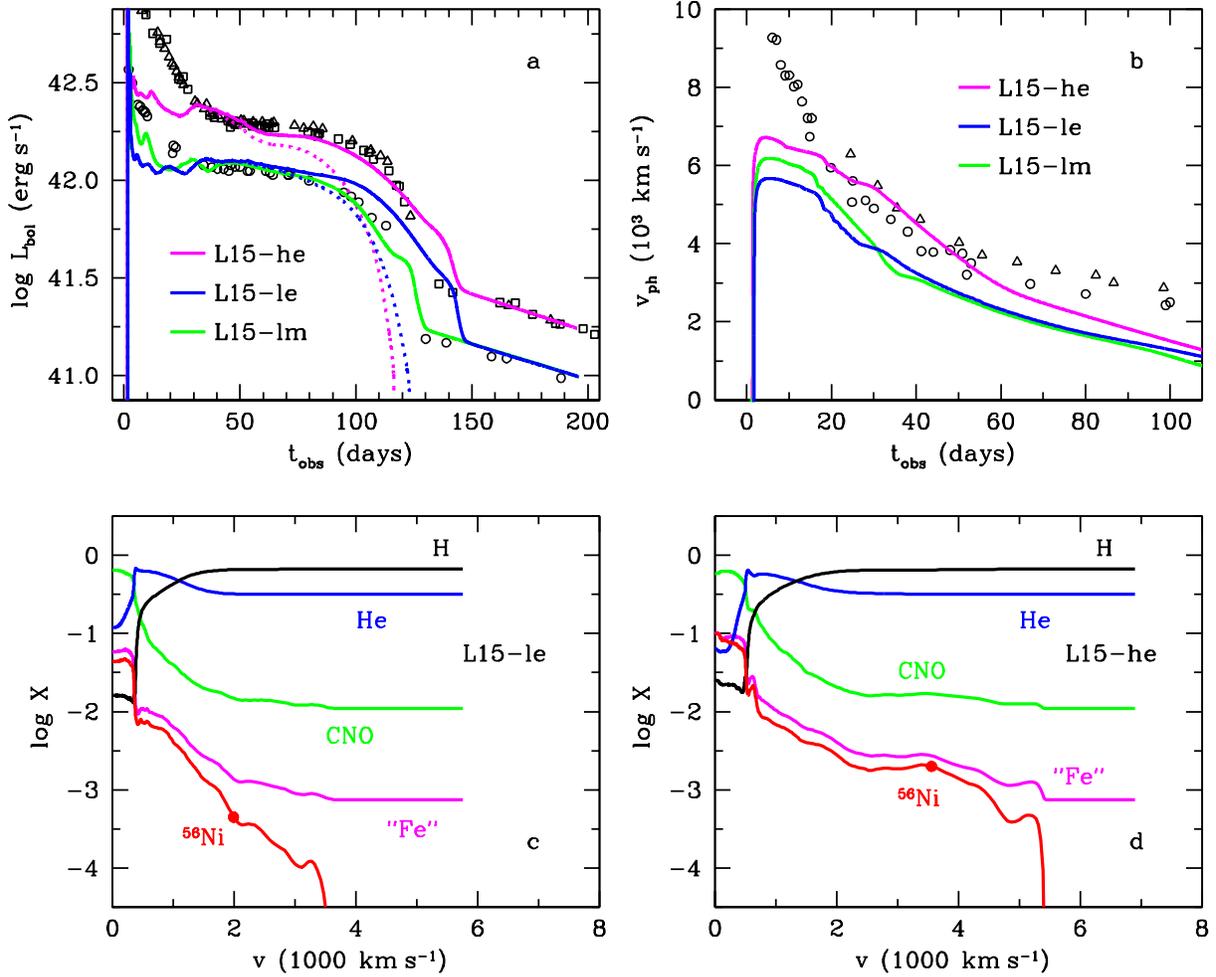}
   \caption{%
   Dependence of the bolometric light curve, the evolution of the photospheric
      velocity, and the chemical composition at $t=50$\,days on the explosion energy.
   Panel \textbf{a} shows the light curves and panel \textbf{b} the evolution
      of the photospheric velocity for models L15-lm (green solid line),
      L15-le (blue solid line), and L15-he (magenta solid line) compared with
      the observations.
   Bolometric light curves of the normal type IIP SN~1999em (open circles)
      and the luminous type IIP SN~2004et (open triangles and squares) are
      estimated from the $UBVRI$ observations of \citet{ECP_03},
      \citet{SASM_06}, and \citet{MDS_10}, respectively.
   The light curves of the corresponding models without radioactive $^{56}$Ni
      are shown by dotted lines.
   Radial velocities at maximum absorption of \ion{Fe}{2} lines are measured
      by \citet{HPM_01} and \citet{LFL_02} for SN~1999em (open circles)
      and by \citet{SASM_06} for SN~2004et (open triangles).
   Panels \textbf{c} and \textbf{d} show the mass fractions of a set of nuclear
      components as functions of velocity for models L15-le and L15-he,
      respectively.
   The red line gives the mass fraction of radioactive $^{56}$Ni, and
      the magenta line represents ``iron-group'' elements containing
      the $\alpha$-nuclei from $^{28}$Si to $^{52}$Fe in the averaged 3D
      explosion models.
   Red bullets mark the outer boundary of the bulk of $^{56}$Ni
      containing $97\%$ of the total $^{56}$Ni mass.
   }
   \label{fig:depeng}
\end{figure*}
Figure~\ref{fig:aphexp} permits us to estimate the time at which the
   hydrodynamic flow approaches homologous expansion.
For the velocity range from 1000 to 4000\,km\,s$^{-1}$, containing a 
   significant mass of radioactive $^{56}$Ni, this time is as large
   as 6\,days, when energy deposition by radioactive decay of $^{56}$Ni
   to $^{56}$Co already becomes important and might affect the dynamics
   of the $^{56}$Ni-rich ejecta, too.
This physical process is not yet included in our current numerical code, but
   it is a subject of future investigation.
Nevertheless, we consider our 3D neutrino-driven simulations mapped at 
   $t_\mathrm{map}$ as an acceptable approximation of the final mixing of
   heavy elements and hydrogen in velocity space, in particular of the inner
   regions, which we focus on for the light-curve discussion in the present
   work.

\subsection{Light curve}
\label{sec:results-lcmod}
%
The development of an ordinary SN~IIP consists of the following basic stages:
   shock breakout, adiabatic cooling phase, phase of cooling and
   recombination wave (CRW), phase of radiative diffusion cooling, and
   a radioactive tail \citep[e.g.,][for details]{Utr_07}.
As an illustrative example, we consider our fiducial model L15-le
   (Figs.~\ref{fig:depeng}a and b).
During shock breakout from the stellar surface and the adiabatic cooling
   phase a narrow peak in the bolometric luminosity forms, and the luminosity
   decreases for the next $\sim$20\,days.
By this epoch, the CRW sets in and cools the ejecta, completely dominating
   the SN luminosity to nearly day 100.
A basic CRW property reads that the higher the ratio of explosion energy
   and ejecta mass is, the higher is the luminosity in the CRW phase, because
   the expanding and cooling ejecta radiate more energy.
This dependence is clearly demonstrated by models L15-le and L15-he
   (Table~\ref{tab:3Dsim}, Fig.~\ref{fig:depeng}a).
The release of internal energy by the CRW is followed by cooling by radiative
   diffusion, which starts in the optically thick expelled envelope at about
   day 100 and ends in the semi-transparent medium around day 140.
After exhaustion of internal energy, the radioactive decay of $^{56}$Ni and
   $^{56}$Co nuclides dominates the luminosity beyond day 150.
Before this moment and during the radioactive tail the bolometric light curve
   depends on the amount of radioactive material and its distribution over
   the ejecta.

Our 3D supernova simulations are characterized by the explosion energy, the
   total amount of radioactive $^{56}$Ni, and the amount of macroscopic mixing
   of $^{56}$Ni and hydrogen-rich matter occurring during the SN explosion.
The total mass of radioactive $^{56}$Ni is higher in model L15-he than 
   in model L15-le because of its larger explosion energy
   (Table~\ref{tab:3Dsim}).
At the same time, these models demonstrate that the larger is the explosion
   energy, the more intense is the mixing of radioactive $^{56}$Ni in velocity
   space, and the larger is the mass fraction of hydrogen in the inner layers
   of the ejecta (Figs.~\ref{fig:depeng}c and d).
We note that the minimum velocity of hydrogen-rich matter is as low as
   zero.

\section{Origin of the luminosity spike}
\label{sec:orgnlsp}
%
\begin{figure}[t]
   \includegraphics[width=\hsize, clip, trim=16 158 70 321]{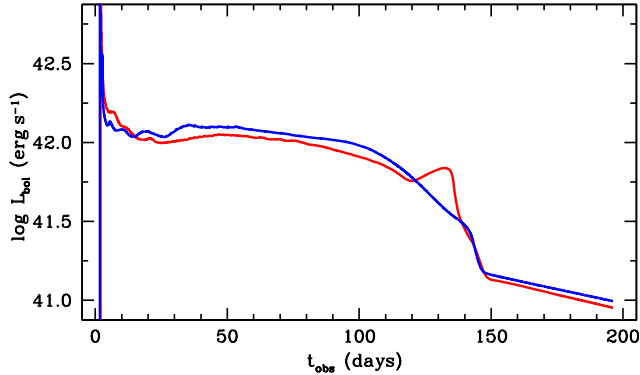}
   \caption{%
   Comparison of the light curves for the averaged 3D neutrino-driven explosion
       model L15-le (blue line) and the 1D piston-driven model L15-pn
       (red line) (Table~\ref{tab:3Dsim}).
   }
   \label{fig:3D1Dlc}
\end{figure}
\begin{figure*}[t]
\centering
   \includegraphics[width=0.45\hsize, clip, trim=23 141 37 99]{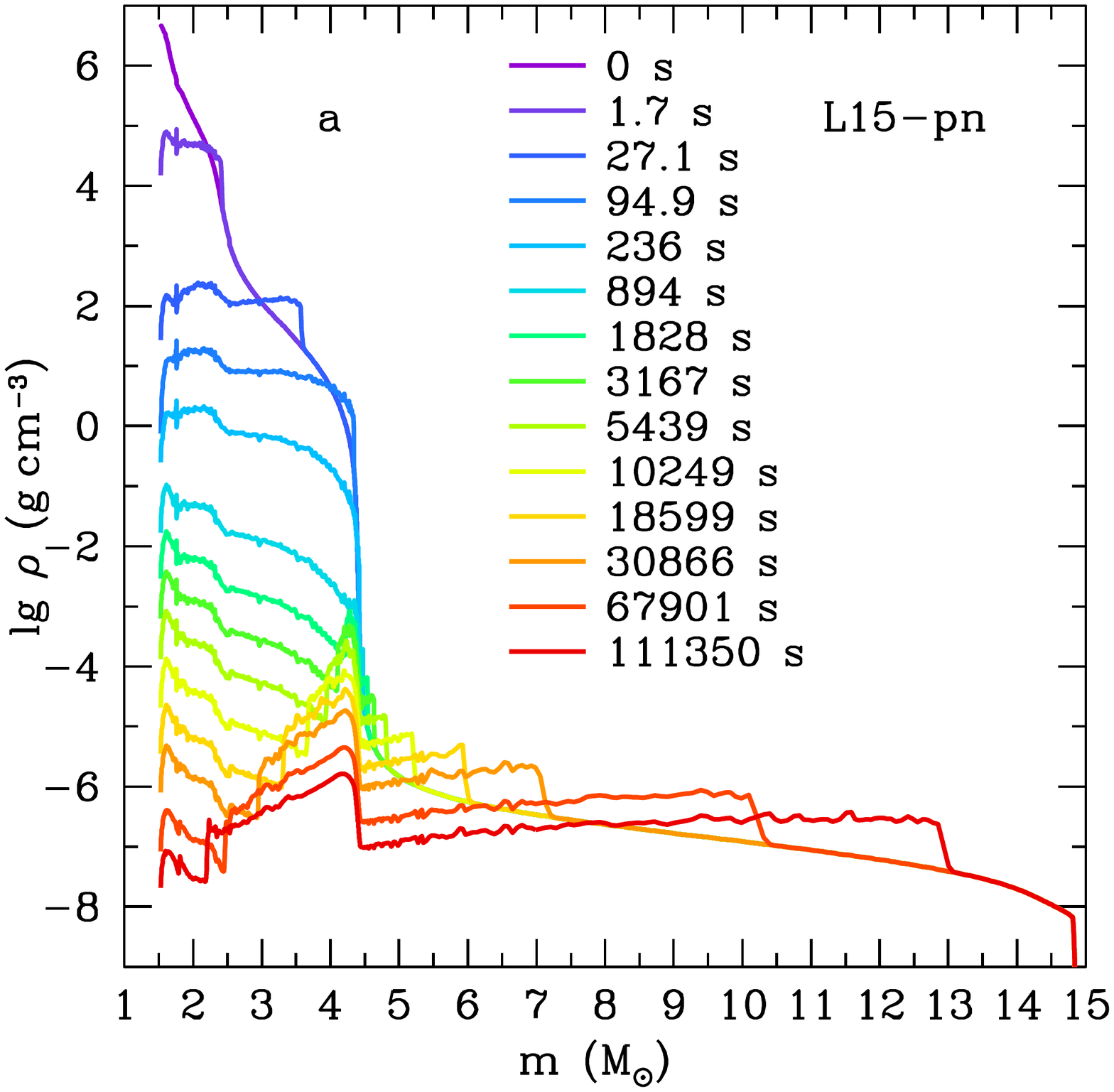}
\hspace{0.25truecm}
   \includegraphics[width=0.45\hsize, clip, trim=23 141 37 99]{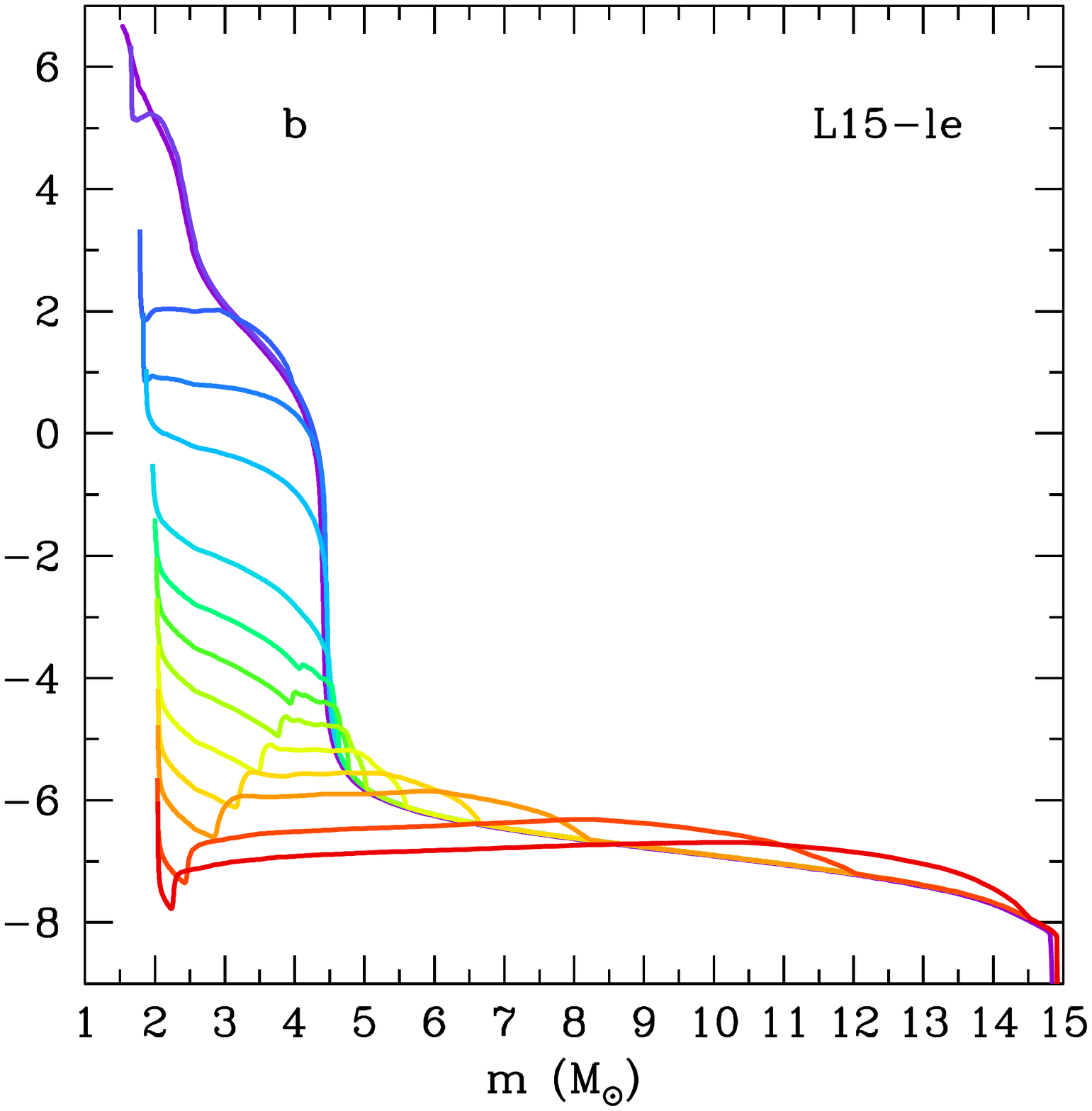}\\
\vspace{0.25truecm}
   \includegraphics[width=0.45\hsize, clip, trim=23 141 37 99]{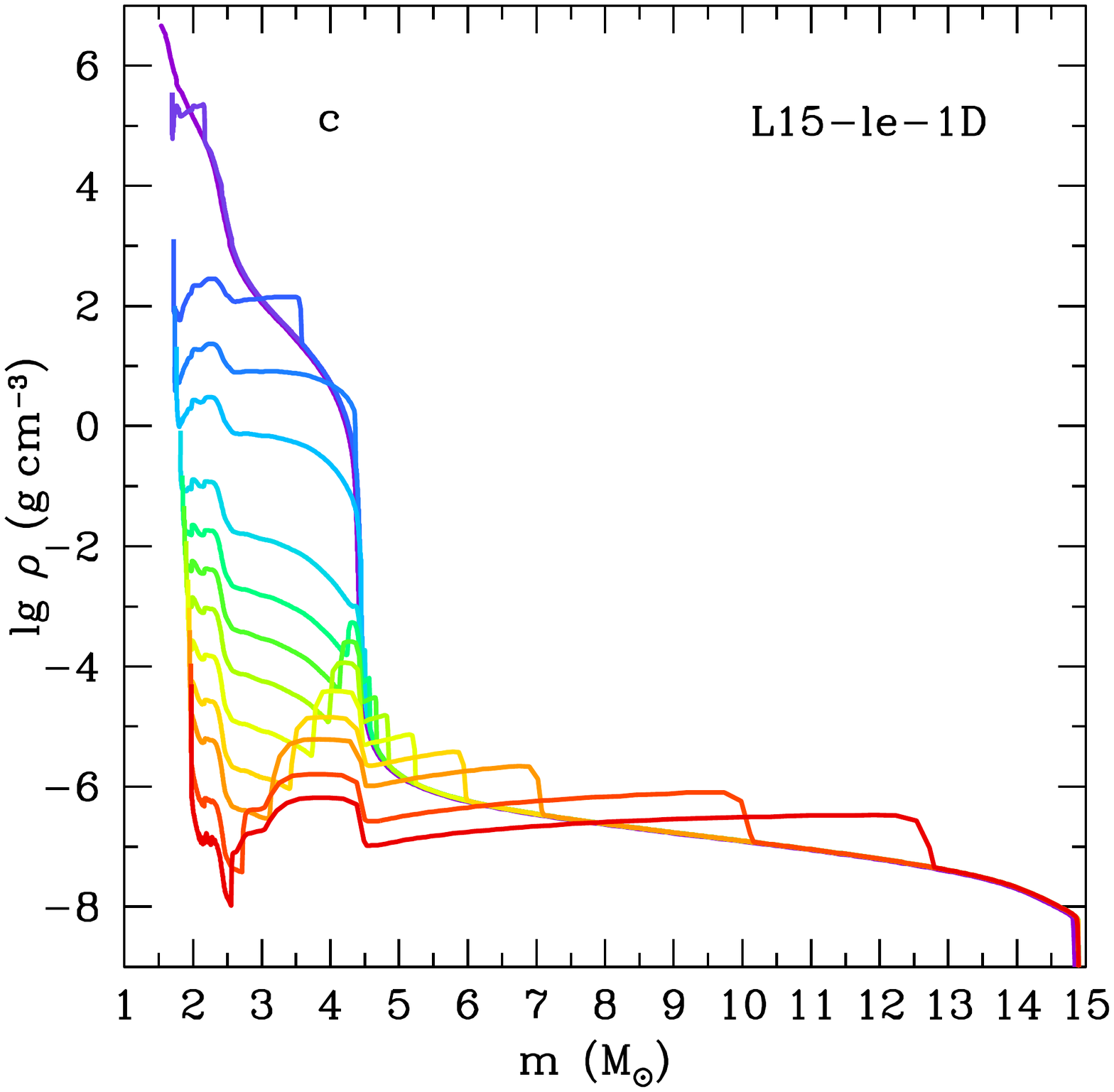}
\hspace{0.25truecm}
   \includegraphics[width=0.45\hsize, clip, trim=23 141 37 99]{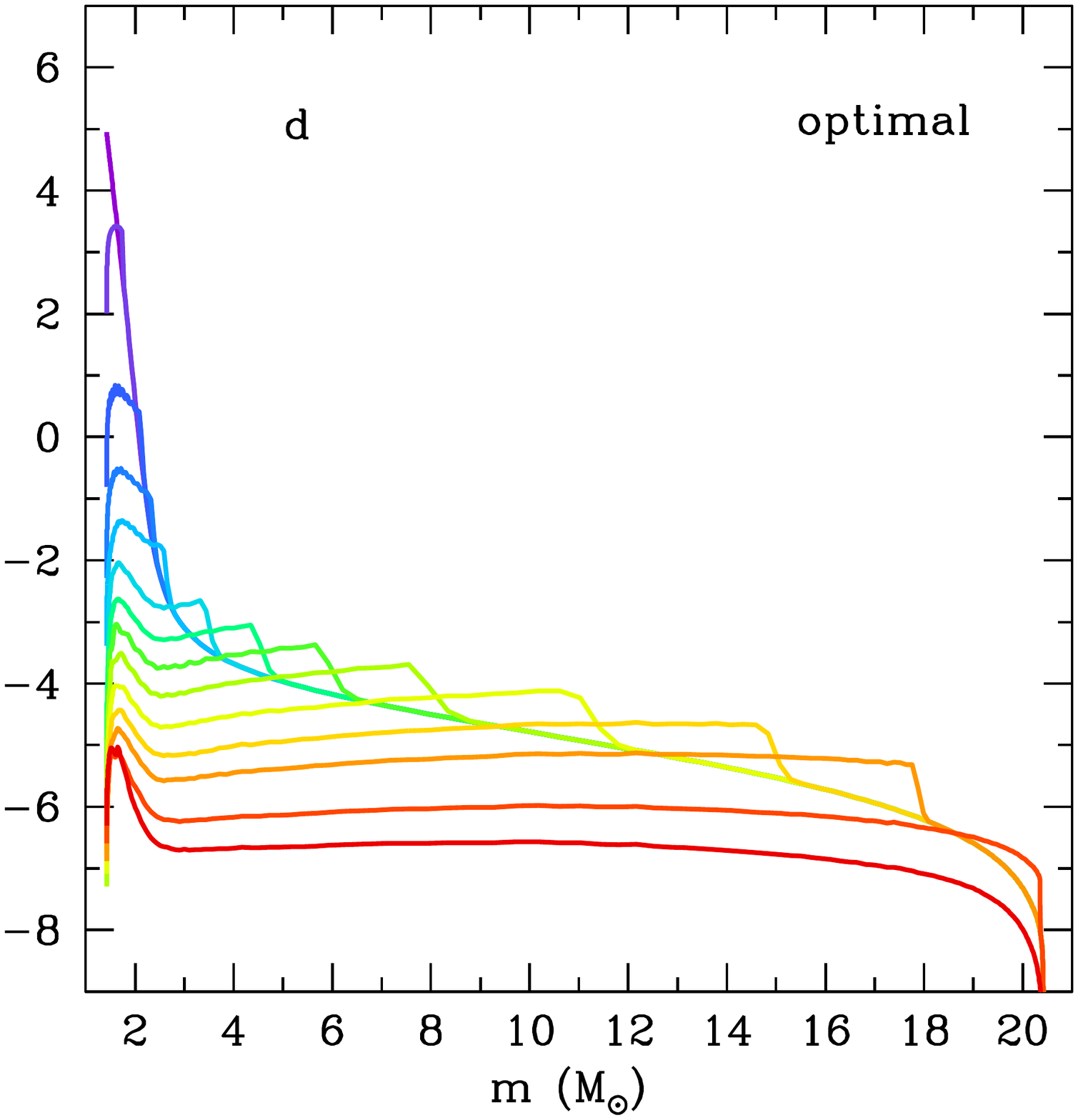}\\
   \caption{%
   Evolution of the density distribution during shock propagation until
      the time of mapping (at $111\,350$\,s) for the 1D piston-driven
      model L15-pn (panel \textbf{a}), the averaged 3D neutrino-driven
      model L15-le (panel \textbf{b}), the 1D neutrino-driven model L15-le-1D
      (panel \textbf{c}), and the 1D optimal model of \citet{Utr_07}
      (panel \textbf{d}).
   The times are measured from core bounce in neutrino-driven
      explosions and from the onset of the explosion for models
      triggered by a piston.
   The density profile at time zero gives the structure of the pre-SN 
      model. 
   Note the appearance of a density step (contact discontinuity) at the outer
      edge of the helium core (at about 4.3\,$M_{\sun}$) after the SN shock
      has crossed the He/H composition interface in models L15-pn and L15-le-1D.
   }
   \label{fig:evlden}
\end{figure*}
As mentioned above, 1D hydrodynamic models based on evolutionary pre-SN models
   and initiated by a piston-driven explosion exhibit an unobserved spike
   in the luminosity decline from the plateau to the radioactive tail.
Our model L15-pn reproduces this general result, while our averaged
   3D neutrino-driven explosion model L15-le and hydrodynamical models
   based on nonevolutionary pre-SN models \citep{Utr_07} show a
   monotonic luminosity decline (see Fig.~\ref{fig:3D1Dlc}).
This situation raises the question what causes the luminosity spike in
   the light curves of hydrodynamic explosion models of evolutionary
   progenitors?

\begin{figure}[t]
   \includegraphics[width=\hsize, clip, trim=16 158 70 321]{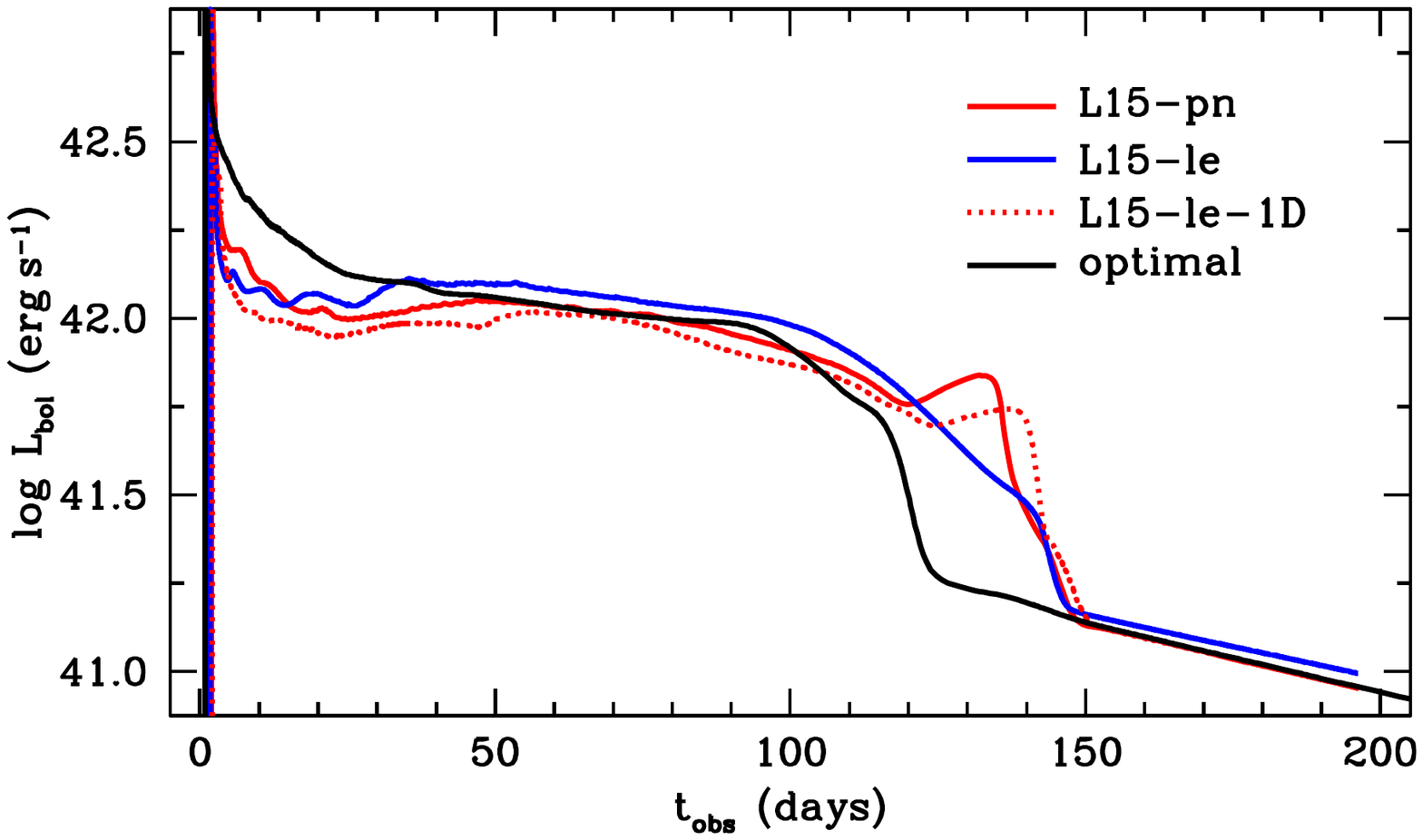}
   \caption{%
   The light curves corresponding to the four models in Fig.~\ref{fig:evlden}:
      model L15-pn (red solid line), model L15-le (blue line), model L15-le-1D
     (red dotted line), and the optimal model of \citet{Utr_07} (black line).
   }
   \label{fig:complc}
\end{figure}

3D neutrino-driven explosion simulations differ from 1D hydrodynamical
   simulations by vigorous radial mixing between the core and the outer
   stellar layers, which significantly modifies both the density distribution
   and the chemical composition in the inner ejecta with respect to
   the spherically symmetric case.
To study the influence of both effects on the luminosity spike, we compare
   the behavior of model L15-pn based on the evolutionary pre-SN model L15
   and exploded by means of a 1D piston (Fig.~\ref{fig:evlden}a), model L15-le
   based on the pre-SN model L15 and a 3D neutrino-driven explosion simulation
   (Fig.~\ref{fig:evlden}b), model L15-le-1D computed as neutrino-driven
   explosion with the {\sc Prometheus} code in 1D and having the somewhat
   lower explosion energy of 0.39\,B compared to 0.54\,B for model L15-le
   (Fig.~\ref{fig:evlden}c), and the optimal model of \citet{Utr_07} based on
   a nonevolutionary pre-SN model and triggered by a 1D piston-driven explosion
   (Fig.~\ref{fig:evlden}d).
The light curves corresponding to these four models are displayed in
   Fig.~\ref{fig:complc}, where the presence or absence of the blip in the
   decline from the plateau to the radioactively powered tail is the most
   relevant feature for the discussion following below.
Because of the lower explosion energy of model L15-le-1D, the plateau of its
   light curve is at a slightly lower level.

In the spherical piston model L15-pn (Fig.~\ref{fig:evlden}a) the shock
   wave crosses the C+O/He and He/H composition interfaces after $t=1.7$\,s
   and $94.9$\,s, respectively.
When it passes the composition interfaces and subsequently propagates 
   into the helium layer and the hydrogen envelope, respectively,
   the shock decelerates and a reverse shock forms.
The shock deceleration causes the formation of a contact discontinuity
   at the He/H composition interface (at about 4.3\,$M_{\sun}$), which is visible as a step in
   the density profile of the last four snapshots ($t \ge 18\,599$\,s).
At the time of mapping ($t=111\,350$\,s), when the main shock is about to
   reach the stellar surface, the density step is very pronounced.

In contrast to model L15-pn, the averaged density distribution of the
   3D model L15-le is nearly flat in the region between the SN
   shock and the reverse shock (once the latter forms at around $t=1828$\,s) 
   without any signature of a density step at the location of the He/H 
   interface (Fig.~\ref{fig:evlden}b).
The flatness of the density distribution reflects a characteristic feature
   of our 3D (neutrino-driven) explosion simulations, namely that large-scale
   macroscopic mixing occurs (see Sect.~\ref{sec:results-3Dexp}) which
   smoothens the density distribution. 

The evolution of the density distribution of the 1D piston-driven model L15-pn
   (Fig.~\ref{fig:evlden}a) simulated with the hydrodynamic code {\sc Crab}
   is similar to that of the 1D neutrino-driven model L15-le-1D
   (Fig.~\ref{fig:evlden}c) simulated with the hydrodynamic code
   {\sc Prometheus}.
When the shock passes the He/H interface, again a reverse shock and subsequently
   a contact discontinuity form, the latter being visible as a density step
   for $t \ge 1828$\,s (Figs.~\ref{fig:evlden}a and c).

Finally, Figure~\ref{fig:evlden}d shows the optimal model of \citet{Utr_07}
   which is based on a nonevolutionary pre-SN model and exploded with
   an 1D piston. 
Its internal structure, which was obtained by comparing observational data
   with adequate hydrodynamic explosion models, is quite different from
   that of the evolutionary pre-SN model L15, showing no sharp density
   gradients which are typical for the pre-SN model L15
   (Figs.~\ref{fig:denmr}a and b).
Hence, the shock propagation does not give rise to any step-like features in 
   its density distribution (Fig.~\ref{fig:evlden}d).
It is interesting to note that the density distributions of the 3D
   neutrino-driven model L15-le and the optimal model exploded by a 1D
   piston are similar at the time of mapping ($t=111\,350$\,s;
   Figs.~\ref{fig:evlden}b and d).

\begin{figure}[t]
   \includegraphics[width=\hsize, clip, trim=17 158 72 317]{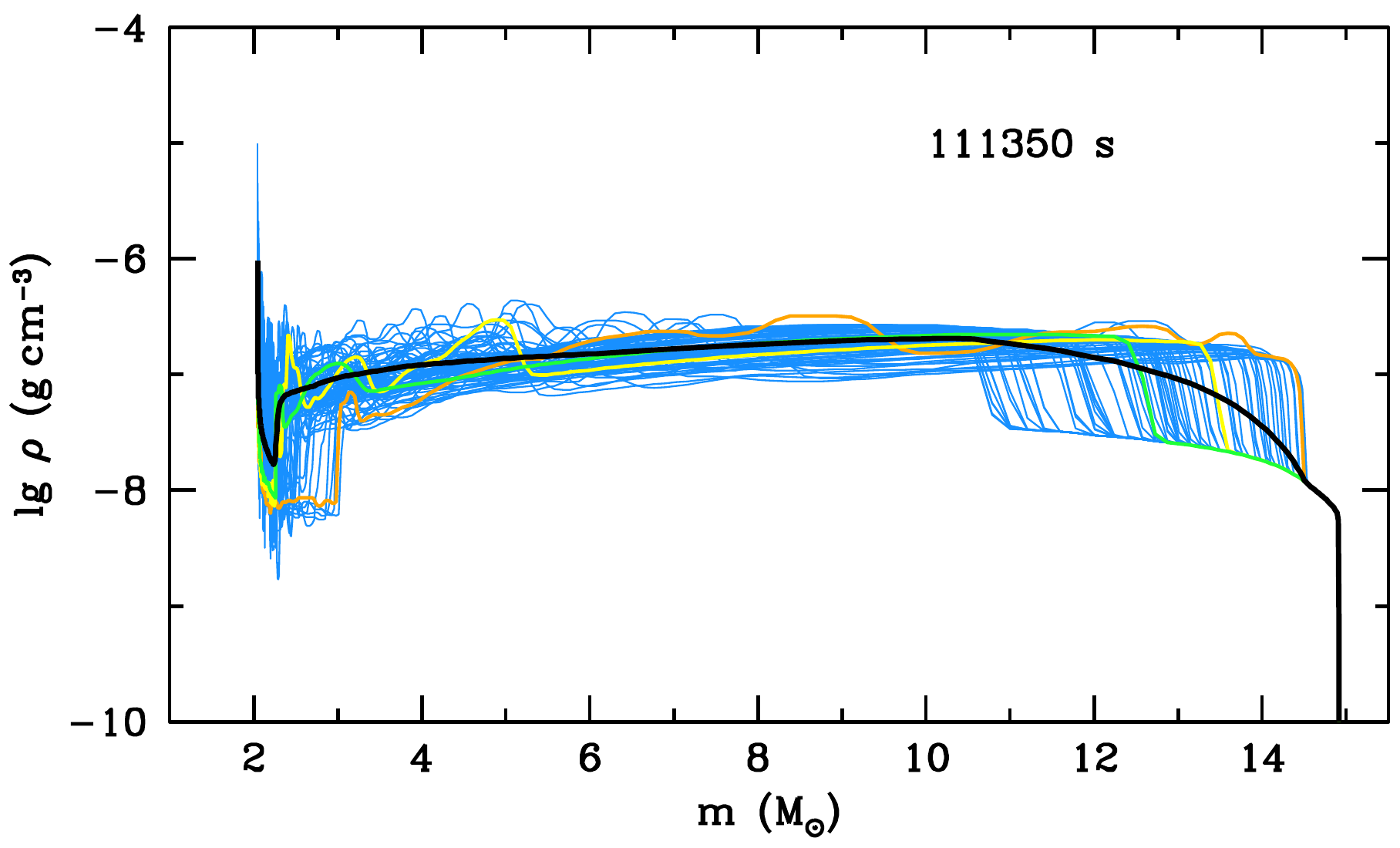}
   \caption{%
   Density as a function of interior mass for model L15-le at the time of
      mapping ($t= 111\,350$\,s).
   The density profiles along different angular directions are sampled
      into bins of 30 degree width (blue lines).
   Three selected profiles are colored in green, yellow, and orange.
   The angular-averaged density profile is shown in black for comparison.
   }
   \label{fig:3dmix}
\end{figure}
To understand the origin of the flat density distribution in the
   averaged 3D neutrino-driven explosion model L15-le, we show in 
   Fig.~\ref{fig:3dmix} density profiles from this model along different 
   angular directions at the time of mapping.
Because the shock is globally deformed in this model, the density profiles
   exhibit jumps at the shock front that are spread out over a mass range
   from 10.5 to 14.5\,$M_{\sun}$ (Fig.~\ref{fig:3dmix}, blue lines). 
Angular averaging transforms this set of sharp fronts into a smooth
   density distribution (black line) which is similar to that in the outer
   $\sim$5\,$M_{\sun}$ in the 1D models (compare Fig.~\ref{fig:evlden}b
   with Figs.~\ref{fig:evlden}a and c).

Because intense turbulent mixing initiated by Rayleigh-Taylor
   instabilities occurs in 3D simulations, the pronounced density steps 
   that are present in 1D models at the (C+O)/He and He/H interfaces are 
   smeared out, too.
To illustrate this point, we marked in Fig.~\ref{fig:3dmix} three 
   profiles along particular angular directions that are characterized 
   by a density step at about 3\,$M_{\sun}$ (green line),
   5\,$M_{\sun}$ (yellow line), and 9\,$M_{\sun}$ (orange line).
Angular averaging of the 3D hydrodynamic flow results in a nearly
   flat density distribution inside the helium core and in the vicinity of
   the (C+O)/He and He/H composition interfaces (Fig.~\ref{fig:3dmix},
   black line) with no reminiscence of any density step.

\begin{figure}[t]
   \includegraphics[width=\hsize, clip, trim=16 158 70 321]{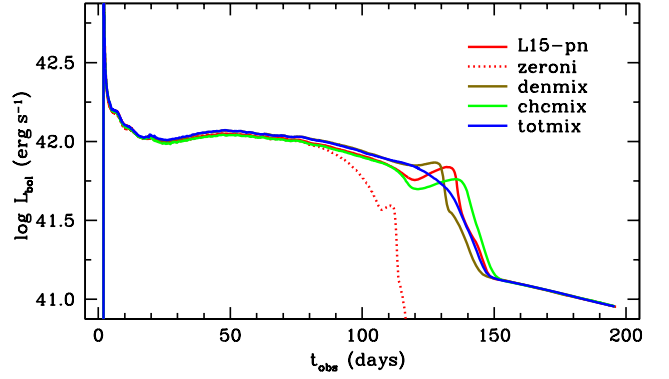}
   \caption{%
 The nature of the luminosity spike in the light curve for the 1D piston-driven
      explosion simulations is illustrated by the reference model L15-pn
      (red solid line) (Table~\ref{tab:3Dsim}) and those constructed on
      its basis.
   The red dotted line is the light curve of model L15-pn without radioactive
      $^{56}$Ni.
   The rest of the models are calculated as L15-pn up to the epoch of
      111\,350\,s and then continued after smoothing of the density step
      (model denmix, olive line), averaging of the chemical composition
      with a boxcar mass width of $1.75\,M_{\sun}$ (model chcmix,
      green line), and both modifications (model totmix, blue line).
   }
   \label{fig:lspike}
\end{figure}

Besides flattening the density distribution, turbulent mixing also causes
   macroscopic mixing of the chemical composition.
This property of hydrodynamic models based on 3D neutrino-driven explosion
   simulations implies that the original chemical composition of pre-SN
   models is unrealistic for modeling light curves.
Moreover, light curves computed from 1D hydrodynamic models which are
   based on evolutionary, unmixed pre-SN models show the mentioned unobserved
   luminosity spike towards the end of the plateau phase, which reflects
   the fact that the photosphere crosses the sharp He/H interface at
   $\sim$4.3\,$M_{\sun}$ in these models, before it enters and moves through
   the helium core.
This sharp composition interface separating hydrogen-rich and helium-rich
   matter (Fig.~\ref{fig:chcom}) of quite different opacity favors the
   formation of the pronounced spike.
In model L15-pn that happens around day 135 (Fig.~\ref{fig:lspike}, 
   red solid line).

It is evident that the luminosity spike may be interpreted as an energy excess
   in the inner layers of the ejecta imprinted on the light curve.
This energy excess can be deposited by either the shock wave or gamma-rays
   from radioactive $^{56}$Ni and $^{56}$Co.
Let us now consider the dependence of the luminosity spike on different factors
   in some more detail.
First, even if there were no radioactive $^{56}$Ni in the ejecta, as assumed
   in model L15-pn, version ``zeroni'', the spike feature is still present
   in the light curve, although less pronounced (Fig.~\ref{fig:lspike},
   red dotted line).
This result shows that the luminosity spike is powered by both gamma-rays
   and the shock wave.
The energy excess is indeed produced by day 100 and deposited by the
   shock wave and gamma-rays around the pronounced density step of model
   L15-pn (Fig.~\ref{fig:evlden}a).
Second, when we artificially smooth the density step at the outer edge of
   the helium core at $111\,350$\,s in model L15-pn, version ``denmix'',
   the luminosity spike does not disappear, but the luminosity still increases
   slightly at the end of the plateau phase (Fig.~\ref{fig:lspike}, olive line).
Third, when we impose an artificial ``boxcar'' averaging \citep[cf.][]{KW_09}
   in model L15-pn, version ``chcmix'', with a boxcar mass width of
   $1.75\,M_{\sun}$ at $111\,350$\,s~%
   \footnote{Mixing, applied to the whole star, is mainly efficient at
      the locations of the (C+O)/He and He/H composition interfaces.},
   but keep the radioactive $^{56}$Ni distribution and the density jump unchanged,
   the spike feature is less luminous and slightly shifts to later times
   (Fig.~\ref{fig:lspike}, green line), which is the result of a higher
   optical depth in the helium core giving rise to a longer diffusion time of
   photons.
Thus, neither a smoothed density distribution nor a mixed chemical composition
   at the outer edge of the helium core erase the unobserved luminosity
   spike.

In order to mimic multidimensional effects in spherically symmetric geometry,
   we recomputed the reference model L15-pn with both artificial smoothing of
   the density step around the outer edge of the helium core and artificial
   mixing of the chemical composition with a boxcar mass width of
   $1.75\,M_{\sun}$ at the He/H interface, both performed at $111\,350$\,s.
We find that the simultaneous action of both modifications in the reference
   model L15-pn, version ``totmix'', prevents the formation of the spike in
   the luminosity decline from the plateau to the radioactive tail
   (Fig.~\ref{fig:lspike}, blue line).
These results imply that both the density step at the outer edge of the helium core
   and the unmixed chemical composition of the evolutionary pre-SN model are
   responsible for the presence of the unobserved spike in the light curve
   computed with hydrodynamic models exploded by a piston in spherically
   symmetric geometry.
Finally, on the basis of our numerical experiments, we can state firmly that
   the monotonic luminosity decline from the plateau to the radioactive tail
   in the ordinary type IIP SN~1999em is a manifestation of macroscopic mixing
   during 3D neutrino-driven explosion.

\section{Comparison with observations}
\label{sec:compobs}
%
A comparison of calculated light curves with the observed light curve
   of SN~1999em during the plateau phase (Fig.~\ref{fig:depeng}a) shows that
   model L15-le (blue solid line) with an explosion energy of 0.54\,B
   reproduces the observed light curve better for the given pre-SN
   model L15 than model L15-he (magenta solid line) with an explosion energy
   of 0.93\,B, whereas the calculated light curve of the latter model
   agrees better with the observed light curve of SN~2004et. 
The agreement between the calculated luminosity of model L15-le and that
   observed for SN~1999em only holds for the plateau phase, and does not
   include the initial luminosity peak observed during the first 
   $\sim$30\,days.
This discrepancy is caused, as in the case of SN~1987A \citep{UWJM_15}, by
   the structure of the outer layers of the pre-SN model, which is
   evidently different from that of the real pre-SN star.
Thus, we focus our discussion mostly on the plateau phase of the light curve, 
   neglecting the initial luminosity peak.

\begin{figure*}[t]
\centering
   \includegraphics[width=0.9\hsize, clip, trim=32 384 72 202]{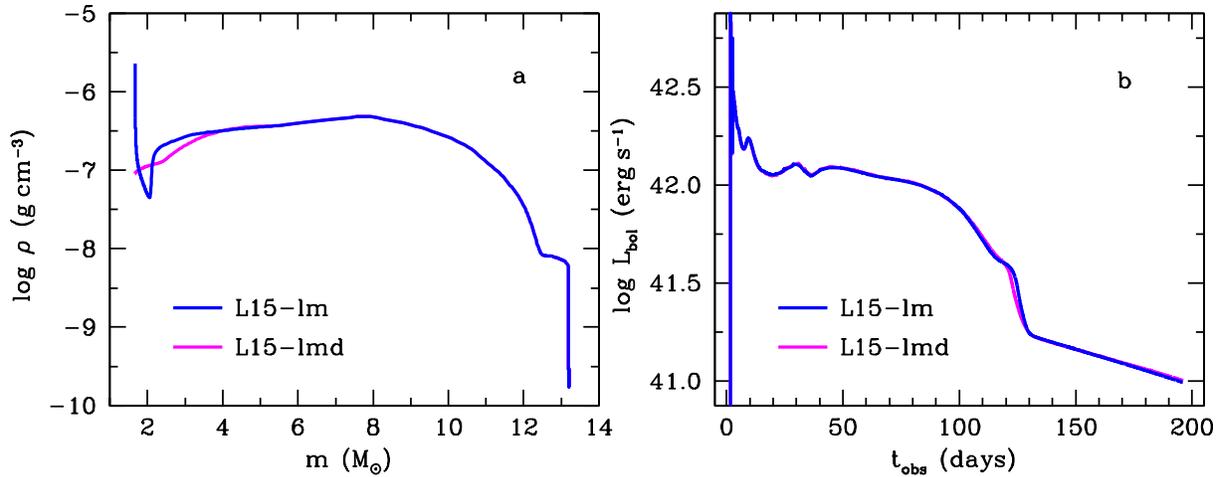}
   \caption{%
   Impact of the density structure in the innermost layers of the ejecta
      at the time of mapping ($t= 111\,350$\,s) (panel~\textbf{a}) on
      the light curve (panel~\textbf{b}) is shown by comparing models L15-lm
      (blue line) and L15-lmd (magenta line).
   }
   \label{fig:denspk}
\end{figure*}
The duration of the plateau phase in the fiducial model L15-le is longer than
   that for SN~1999em (Fig.~\ref{fig:depeng}a).
Because it is well known that the duration of the plateau phase depends mainly
   on the ejecta mass \citep[e.g.,][]{Utr_07}, it is not surprising that
   model L15-lm, which has the same explosion energy as model L15-le but
   an ejecta mass reduced according to the mass-loss estimate by
   \citet{CCU_07}, fits the observed light curve better (except for
   the initial luminosity peak, which gets closer to the observations but
   is still deficient).
However, the reduced ejecta mass of model L15-lm does not remove a shoulder-like
   feature present in the computed light curves during the luminosity
   decline from the plateau to the radioactive tail (Fig.~\ref{fig:depeng}a).
As mentioned above, this feature is not seen in the observations of ordinary
   SNe IIP (Fig.~\ref{fig:obslc}).
The decline occurs when the ejecta become semi-transparent for photons
   (optical depth of order unity), and the SN luminosity forms in the
   innermost layers of the ejecta.
The latter are characterized by macroscopic inhomogeneities in their chemical
   composition and in the density distribution in the mixing zone.
These inhomogeneities can reduce the effective opacity compared to the
   homogeneous case, and consequently affect the luminosity.
Using model L15-lm, we explored this clumping effect on both photon and
   gamma-ray transport in the framework of the approach developed by
   \citet{UC_15}.
We found that the shoulder feature is insensitive to clumping because the
   optical depth of clumps does not exceed unity during the relevant phase.

At the mapping epoch, the density profiles of the averaged 3D models show
   a structural feature in the innermost layers of the ejecta at a mass
   coordinate of $\approx$2\,$M_{\sun}$ (Fig.~\ref{fig:evlden}b).
To study the influence of this feature, we performed an additional simulation,
   model L15-lmd, which is based on model L15-lm and in which we
   artificially flattened the density distribution in the central region
   at the mapping epoch (Fig.~\ref{fig:denspk}a).
We find that in the resulting light curve the shoulder-like feature at
   the luminosity decline from the plateau to the radioactive tail is
   less pronounced, but still visible (Fig.~\ref{fig:denspk}b).

The existence of a shoulder-like feature in the computed light curve during
   the luminosity decline from the plateau to the radioactive tail in contrast
   to the observed SN light curve might be attributed to structural differences
   in the progenitor star compared to the model L15 that we employed.
Alternatively, it might point to still missing effects in our treatment of
   the SN explosion, for example to mixing induced by $^{56}$Ni-decay heating,
   which could be accounted for only by 3D simulations continued to much later
   times.

A good fit of the calculated bolometric light curve of models L15-le and
   L15-lm to the observations of SN~1999em during the plateau phase does not ensure
   that these models also give the correct evolution of the photospheric
   velocity (Figs.~\ref{fig:depeng}a and b).
Actually, the disagreement between the calculated evolution of the photospheric
   velocity and the observed radial velocities of spectral lines during the
   first $\sim$20\,days is serious and casts doubts on a perfectly proper
   choice of the pre-SN model.
At later epochs the disagreement is less, but the computed photospheric
   velocities are well below the observed ones.
With its higher explosion energy the velocities of model L15-he agree better
   with the observations of SN~1999em between day 20 and day 60, but its
   bolometric luminosity significantly exceeds the observed one.

Evaluating the total mass of radioactive $^{56}$Ni by equating the observed
   bolometric luminosity in the radioactive tail to the gamma-ray luminosity
   gives a mass of $\approx$0.036\,$M_{\sun}$ for SN~1999em, which falls
   in between the minimum, $M_\mathrm{Ni}^{\,\mathrm{min}}$, and maximum,
   $M_\mathrm{Ni}^{\,\mathrm{max}}$, values obtained with our models
   (Table~\ref{tab:3Dsim}).
Thus, our 3D neutrino-driven simulations are able to synthesize the required
   amount of ejected radioactive $^{56}$Ni.
Note that hydrodynamic models without radioactive $^{56}$Ni in the ejecta
   provide a lower bound on the plateau phase duration for a given pre-SN
   model and explosion energy (Fig.~\ref{fig:depeng}a, dotted lines).

An analysis of SN~1999em observations shows that the distribution
   of the bulk of $^{56}$Ni can be approximated by a sphere with a velocity
   of 1500\,km\,s$^{-1}$ that is shifted towards the far hemisphere by
   about 400\,km\,s$^{-1}$ \citep{ECP_03}.
As already mentioned in Sect.~\ref{sec:results-3Dexp}, the final morphology
   of the $^{56}$Ni-rich ejecta in velocity space possesses a strong
   dipole component in our models, which is characterized by a velocity
   asymmetry (with respect to the motion of the center of mass of $^{56}$Ni)
   of nearly 300\,km\,s$^{-1}$ for model L15-le, and thus comparable with
   the observed shift.
Only model L15-le yields a maximum velocity of the bulk mass of $^{56}$Ni
   of $\sim$2000\,km\,s$^{-1}$ consistent with the observations, while
   model L15-he with an explosion energy of 0.93\,B produces mixing that is
   too strong (Figs.~\ref{fig:depeng}c and d).
Observational evidence for the existence of hydrogen-rich matter within
   the core of heavy elements of SN~1999em implies a deep macroscopic mixing
   down to zero velocity \citep{MJS_12}.
It is remarkable that all of our 3D neutrino-driven simulations show such a deep
   mixing of hydrogen-rich matter (Figs.~\ref{fig:depeng}c and d).

Polarimetric data of SN~1999em presented by \citet{LFAB_01} show that
   continuum polarization is $\approx$0.2$\%$ on day 7 and $\approx$0.3$\%$
   on days 40 and 49.
For our 3D simulations of model L15-le, we approximated the asphericity of
   the outer layers by an ellipsoidal shape and found in
   Sect.~\ref{sec:results-3Dexp} that the maximum ratio of the semiaxes is
   $\approx$1.15. 
According to \citet{Hof_91}, this ratio results in a linear polarization of
   $\approx$0.7$\%$ for an inclination angle of $90^{\circ}$.
Analyzing the double-peak profile of H$\alpha$ at the nebular epoch in terms
   of an asymmetric bipolar configuration for radioactive $^{56}$Ni,
   \citet{Chu_07} argued that this bipolar configuration should be oriented
   at an inclination angle of $39^{\circ}$ to match the observations.
Our 3D neutrino-driven simulations of model L15-le revealed the fact that
   a direction of the strong dipole component of $^{56}$Ni-rich matter nearly
   coincides with that of a larger deformation of the main shock around its
   breakout.
The reason behind is that a larger deformation occurs in the direction of
   a stronger explosion which, in turn, leads to a larger production of
   $^{56}$Ni.
A more accurate estimation of the angle between the directions discussed
   gives a value of about $10^{\circ}$.
Given the relations computed by \citet{Hof_91}, an uncertainty of $10^{\circ}$
   in the inclination angle for an ellipsoidal surface of the photosphere
   reduces the polarization estimated above to $\approx$0.3$\pm$0.1$\%$,
   which is consistent with the broadband polarimetry of SN~1999em.

Note that the disagreement between the calculated photospheric velocity and
   the observed one during the first 40\,days (Fig.~\ref{fig:cmpobs}b) cannot
   be explained by a viewing-angle effect in a fully 3D radiation transport
   calculation for the neutrino-driven explosion model L15-le.
The ratio of observed to calculated photospheric velocity during the discussed
   period of evolution is as large as two, which is significantly greater than
   the maximum ratio of the semiaxes of the ellipsoid of about 1.15.
In other words, a viewing angle effect should not exceed about 15$\%$ in
   the photospheric velocity.

\section{Discussion and conclusions}
\label{sec:discssn}
%
\begin{figure*}[t]
\centering
   \includegraphics[width=0.9\hsize, clip, trim=32 384 72 202]{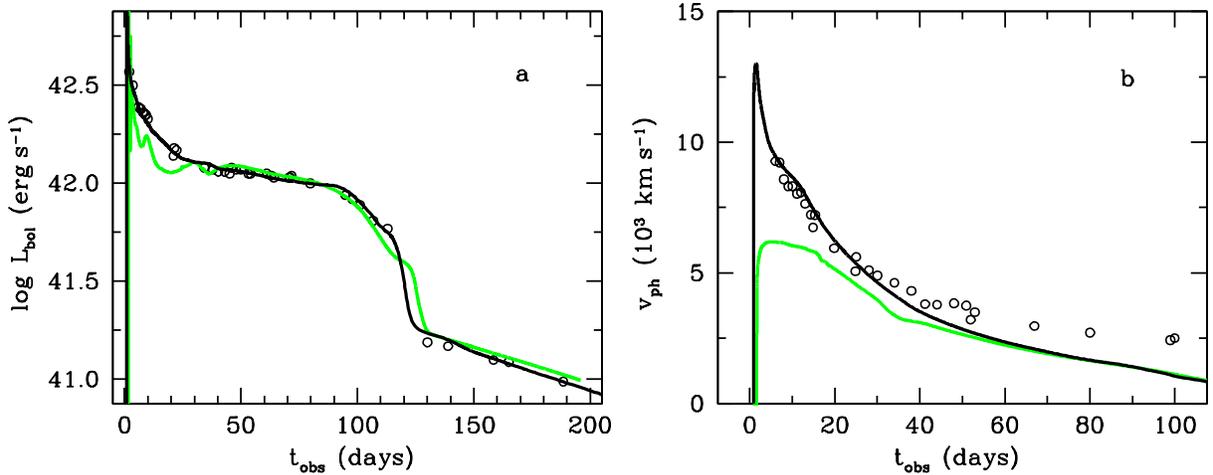}
   \caption{%
   Bolometric light curve (panel \textbf{a}) and photospheric velocity
      (panel \textbf{b}) of model L15-lm (green lines) compared with
      the corresponding observations of SN~1999em (open circles)
      (see Fig.~\ref{fig:depeng}).
   For a comparison, the black lines show the bolometric light curve and the
      photospheric velocity of an optimal model with a nonevolutionary pre-SN
      structure \citep[][see also Fig.~\ref{fig:denmr}]{Utr_07}.
   }
   \label{fig:cmpobs}
\end{figure*}
The present paper is our second attempt to model the light curves of
   type IIP supernova explosions based on 3D explosion models
   (our first one was concerned with the peculiar type IIP SN~1987A
   \citet{UWJM_15}).
We find that 3D neutrino-driven explosion simulations based on the
   evolutionary pre-SN model L15 of \citet{LSC_00} with an explosion energy
   around 0.5\,B are able to reproduce the
   overall behavior of the bolometric light curve of SN~1999em reasonably
   well, along with the production of radioactive $^{56}$Ni and mixing of
   hydrogen deep into the ejecta.
However, the luminosity and the photospheric velocity during the first
   40\,days are inconsistent with observations of SN~1999em
   (Fig.~\ref{fig:cmpobs}).
On the other hand, \citet{Utr_07} constructed an optimal hydrodynamic model
   of the explosion of SN~1999em based on a comparison with detailed
   observational data.
This optimal model has an initial radius of $500\,R_{\sun}$, an ejecta mass of
   $19\,M_{\sun}$, an explosion energy of 1.3\,B, and a total $^{56}$Ni mass
   of $0.036\,M_{\sun}$ (Tables~\ref{tab:presnm} and \ref{tab:3Dsim}).
The density profile of the optimal, nonevolutionary pre-SN star is compared
   with that of the evolutionary model L15 in Fig.~\ref{fig:denmr}.

The calculated light curve and photospheric velocity of the optimal model
   agree well with the observational data for the first 40\,days
   (Fig.~\ref{fig:cmpobs}).
This fact confirms our conclusion from Sect.~\ref{sec:compobs} that the
   serious disagreement of light curve models based on evolutionary pre-SN
   models and 3D neutrino-driven simulations with the observations of
   SN~1999em is caused by the inappropriate structure of the outer layers of
   available pre-SN models.
In particular, the luminosity and photospheric velocity are too low in model
   L15-lm during the first 40\,days compared to the observations, which implies
   that the internal energy deposited and the kinetic energy of the outer
   layers should be larger, i.e.\ the explosion energy should be higher than
   in our reference case of neutrino-driven explosion models.
As a measure of this inconsistency we may consider the kinetic energy of
   the outer layers, which power the luminosity during the first 40\,days
   and where extra mass accounts for the considerably higher ejecta mass
   of the optimal explosion model (Table~\ref{tab:3Dsim}).
This kinetic energy amounts to $\approx$40\% of the total kinetic energy of
   the optimal model or about 0.52\,B, which is almost the {\it explosion}
   energy of model L15-lm (0.54\,B), and thus it explains the significant
   difference in the explosion energies of the optimal model (1.30\,B) and
   model L15-lm.
In addition, it is evident from Fig.~\ref{fig:denmr} that the density
   distribution of the outer ejecta of at least 2.5\,$M_{\sun}$ in the
   hydrogen envelope of the real pre-SN RSG star was steeper than in
   the evolutionary model.
Moreover, as pointed out by \citet{UC_08}, turbulent mixing during the 
   explosion should also flatten the jumps in density and chemical composition
   at the Si/O, (C+O)/He, and He/H interfaces. 
The general similarity between the density profiles of the averaged 3D explosion
   model L15-le and the optimal model (Figs.~\ref{fig:evlden}b and d), 
   and between the chemical composition mixed by realistic turbulent mixing
   in 3D models (Figs.~\ref{fig:depeng}c and d) and artificially in the
   optimal model \citep[][Fig.~2]{Utr_07} confirms both assumptions.

It is noteworthy that the macroscopic mixing of $^{56}$Ni and hydrogen-rich
   matter that occurs during the SN explosions of RSG and BSG progenitors
   has different consequences in the corresponding ordinary and peculiar
   SNe IIP.
In the ordinary type IIP SN~1999em, mixing induced by the 3D neutrino-driven
   explosion causes the monotonic luminosity decline from the plateau to the
   radioactive tail (Sect.~\ref{sec:compobs}), while in the peculiar type IIP
   SN~1987A it results in a broader width of the dome-like maximum of
   the light curve \citep{UWJM_15}.
Neglecting the influence of turbulent mixing on the light curve
   produces conspicuous features that are not seen in observations.
It results in a luminosity spike in the case of SN~1999em
   (Sect.~\ref{sec:orgnlsp}) and a half-truncated maximum of the SN~1987A
   light curve \citep{Woo_88}.

Of particular importance is the morphology of the $^{56}$Ni-rich ejecta
   in velocity space, which reveals asymmetry with a strong dipole component
   (Fig.~\ref{fig:3D_models}) consistent with the observations of the 
   ordinary type IIP SN~1999em \citep{ECP_03}.
Such a non-spherical morphology of the $^{56}$Ni-rich ejecta is not
   an exceptional phenomenon of ordinary SNe IIP.
Analyzing the spectroscopic observations of SN~2004dj, \citet{CFS_05} showed
   that the H$\alpha$ line profile and its evolution are well reproduced
   in a model with asymmetric, dipolar $^{56}$Ni ejecta.
\citet{LFG_06} confirmed that $^{56}$Ni was ejected in a non-spherical manner
   in the explosion of SN~2004dj.
Moreover, the asymmetric shape of the H$\alpha$ line in the nebular phase of
   SN~2013ej may be interpreted in terms of asymmetrically ejected $^{56}$Ni
   as well \citep{HWZ_15}.

For ordinary SNe IIP the asymmetric morphology of the $^{56}$Ni-rich ejecta
   affects the bolometric light curve after the plateau phase, especially
   during the luminosity decline from the plateau to the radioactive tail.
The dipolar configuration of the $^{56}$Ni-rich ejecta in our 3D
   neutrino-driven simulations evidently gives rise to viewing angle
   effects for the light curve, which have to be calculated with a 3D
   radiation hydrodynamics solver.
Because of a lack of the latter, this only leaves the possibility of
   discussing the possible influence of 3D radiation transfer on the
   light curve.

An asymmetry with a strong dipole component may be approximated by
   an aspherical explosion which is strongest and leads to most intense mixing
   along one direction, and is weakest and leads to least intense mixing in
   the opposite direction.
If we are oriented along the direction of the strongest explosion, we observe
   an increased luminosity at the end of the plateau phase and a shorter 
   duration of the plateau compared to what a model would show that is
   based on angular-averaged $^{56}$Ni-rich ejecta \citep{Utr_07}.
Observations along the opposite direction would reveal the opposite effect:
   a decreased luminosity at the end of the plateau phase and a longer duration
   of the plateau.
The greater the difference in the extent of $^{56}$Ni mixing is in different
   directions, the greater is the difference between the corresponding light
   curves.
The influence of 3D radiation transfer on the light curve and the viewing angle
   effects are discussed in more detail by \citet{UWJM_15}.

\citet{DGH_14} suggest a method for metallicity determinations based on
   quantitative spectroscopy of SNe IIP during the plateau, in particular
   with oxygen lines.
As pointed out above, these objects are characterized by a wide variety of
   properties and imply a very different extent of matter mixing in velocity
   space or in mass coordinate.
The detailed discussion of SNe IIP as a metallicity probe has to be carried
   out in the context of the corresponding 3D simulations.
In this paper, we study the issue of turbulent mixing in the normal type IIP
   SN~1999em and are only able to give an educated guess how representative
   this case is for the mixing in other SNe IIP.
The inner layers, which are enriched by oxygen by a factor of two compared to
   the outer layers, move at velocities of 1340 and 1880\,km\,s$^{-1}$
   in models L15-le and L15-he, respectively.
Accordingly, oxygen lines can reflect a change in the oxygen content after
   days 97 and 89, respectively, well after the end of the plateau phase.
In these cases, the method of \citeauthor{DGH_14} remains applicable at least
   with respect to oxygen lines.
As to the newly synthesized metals of the iron group, including radioactive
   $^{56}$Ni, we can say that they are mixed up to about 3740 and
   5460\,km\,s$^{-1}$ in models L15-le and L15-he and that the corresponding
   layers of the ejecta enriched by them become visible after days 34 and 30,
   respectively.
From these epochs on we could expect to observe effects of non-thermal
   ionization and excitation associated with radioactive decays in the spectra.

It is instructive to compare two extensive studies of SN~1999em performed by
   \citet{Utr_07} and \citet{BBH_11} on the basis of nonevolutionary
   pre-SN models.
Their favorite hydrodynamic models have comparable ejecta masses, pre-SN radii,
   explosion energies, and total $^{56}$Ni masses, while the extent of outward
   $^{56}$Ni mixing is up to about 660 and 2300\,km\,s$^{-1}$ in velocity
   space, respectively.
This kind of disparity between two hydrodynamic models, calculated with
   quite different radiation hydrodynamics codes, is formally admitted
   because hydrodynamic modeling itself belongs to the class of ill-posed
   inverse problems that lack a unique solution.
However, the analysis of the H$\alpha$ and He\,{\sc i} 10\,830~\AA\ lines
   in SN~1999em during the nebular epoch implies that the $^{56}$Ni
   distribution can be approximated by a sphere with a velocity of
   1500\,km\,s$^{-1}$ \citep{ECP_03}.
Thus, the extent of $^{56}$Ni mixing in the model of \citet{BBH_11} for
   SN~1999em seems to be inconsistent with the spectral observations and
   becomes, as we will see below, critical for their hydrodynamic model.

To simulate the SN~1999em outburst, both \citet{Utr_07} and \citet{BBH_11}
   used nonevolutionary pre-SN models with an artificially mixed
   chemical composition.
Light curves computed from hydrodynamic models with such pre-SN models, which
   have no well-defined helium core (neither in density nor in chemical
   composition) do not exhibit any bump feature when the ejecta contain
   no radioactive $^{56}$Ni (\citealt[Fig.~17a]{Utr_07};
   \citealt[Fig.~12]{BBH_11}).
The corresponding light curve of \citeauthor{Utr_07} fits the observed one
   during the plateau phase, while the light curve of \citeauthor{BBH_11}
   declines faster than observed.
Adding $^{56}$Ni to the model of \citeauthor{Utr_07} does not destroy the
   monotonicity of the light curve.
In contrast, the $^{56}$Ni mixed up to about 700\,km\,s$^{-1}$ in the ejecta
   of the model of \citeauthor{BBH_11} gives rise to a bump feature at
   the end of the plateau, and as a consequence, causes a local minimum in
   the plateau at about day 75.
To compensate this minimum in luminosity and to obtain a nearly flat plateau,
   \citet{BBH_11} invoked an extended $^{56}$Ni mixing up to about
   2300\,km\,s$^{-1}$, which affects the light curve starting from about
   day 35.
We can state that the origin of the bump feature in the light curve of
   the model of \citet{BBH_11} is related to an inadequate pre-SN structure
   and depends on the extent of $^{56}$Ni mixing.
Such a behavior of this bump feature has nothing in common with the origin
   of the luminosity spike in the light curves of hydrodynamic explosion models
   of evolutionary progenitors as discussed in our work.

Comparing results of 3D neutrino-driven explosion simulations and light curve
   modeling with the observations of the ordinary type IIP SN~1999em, we draw
   the following conclusions based on the considered pre-SN model:
\begin{itemize}
\item
3D neutrino-driven explosion simulations reproduce basic properties of
   the overall behavior of the bolometric light curve of SN~1999em along with
   the radioactive $^{56}$Ni production and extent of hydrogen mixing.
There is a pronounced deficit only in the luminosity and the photospheric
   velocity during the first 40\,days compared to the observations of
   SN~1999em.
This shortcoming is caused by the pre-SN structure of the outer stellar layers,
   which is inadequate to match the observed light curve and the evolution of
   the photospheric velocity.
\item
Spectroscopic observations of SN~1999em show that the bulk of the radioactive
   $^{56}$Ni is shifted towards the far hemisphere by about 400\,km\,s$^{-1}$.
This is comparable to the results of our 3D neutrino-driven explosion,
   in which the morphology of $^{56}$Ni-rich matter has an asymmetry with
   a strong dipole component in velocity space with a characteristic shift
   of nearly 300\,km\,s$^{-1}$.
\item
1D piston-driven explosions of evolutionary pre-SN models inevitably
   produce a pronounced spike in the luminosity decline from the plateau
   to the radioactive tail, which disappears only in the framework of
   the 3D neutrino-driven explosion simulations.
Thus, the monotonic luminosity decline from the plateau to the radioactive
   tail observed for ordinary SNe IIP is a manifestation of intense
   turbulent mixing at the He/H composition interface.
\end{itemize}
In summary, we can explain the basic observational data of the ordinary
   type IIP SN~1999em, except for those related to the detailed pre-SN
   structure of the outer layers, within the paradigm of the neutrino-driven
   explosion mechanism.

\acknowledgements
%
We would like to thank Marco Limongi for providing us with the pre-SN data.
V.P.U. was supported by the guest program of the Max-Planck-Institut f\"ur
   Astrophysik.
At Garching, funding by the Deutsche Forschungsgemeinschaft through grant
   EXC 153 ``Origin and Structure of the Universe'' and by the EU through
   ERC-AdG No.\ 341157-COCO2CASA is acknowledged.
Computation of the 3D models and postprocessing of the data were done on Hydra
   at the Rechenzentrum Garching.



\end{document}